\documentstyle[titlepage,12pt]{article}
\begin{document}
\renewcommand{\theequation}{\thesection.\arabic{equation}}
\title{Strings Propagating
in the 2+1 Dimensional Black Hole Anti de Sitter
Spacetime}
\author{A.L. Larsen\thanks{Observatoire de Paris,
DEMIRM. Laboratoire Associ\'{e} au CNRS
UA 336, Ob- \hspace*{6mm}servatoire de Paris et
\'{E}cole Normale Sup\'{e}rieure. 61, Avenue
de l'Observatoire, \hspace*{6mm}75014 Paris, France.}$^,$
\thanks{On leave of absence from NORDITA,
Blegdamsvej 17, Dk-2100 Copenhagen, \hspace*{6mm}Denmark.}
and
N. S\'{a}nchez$^1$}
\maketitle
\begin{abstract}
We study the string propagation in the 2+1 black hole anti de Sitter
background (2+1 BH-ADS). We find the first and second order fluctuations around
the string center of mass and obtain the expression for the string mass. The
string motion is stable, all fluctuations oscillate with real frequencies
and are bounded, even at $r=0.$ We compare with the string motion in the
ordinary black hole anti de Sitter spacetime, and in the black string
background, where string instabilities develop and the fluctuations blow up
at $r=0.$ We
find the exact general solution for the circular string motion in all
these backgrounds, it is given closely and completely in terms of elliptic
functions. For the non-rotating black hole backgrounds the circular strings
have a maximal bounded size $r_m,$ they contract and collapse into $r=0.$ No
indefinitely growing strings, neither multi-string solutions are present in
these backgrounds. In rotating spacetimes, both the 2+1 BH-ADS and the
ordinary Kerr-ADS, the presence of angular momentum prevents the string
from collapsing into $r=0.$
The circular string motion is also completely solved in
the black hole de Sitter spacetime and in the black string background (dual
of the 2+1 BH-ADS spacetime), in which expanding unbounded strings and
multi-string solutions appear.
\end{abstract}
\section{Introduction and Results}
The study of string dynamics in curved spacetime and its associated physical
phenomena that was
started in Refs.\cite{veg1,san1}, has received a systematic and
increasing attention. Approximative [1-4]
and exact [5-9] solving
methods have been developed. Classical and quantum string dynamics have been
investigated in black hole backgrounds \cite{san2,lou}, cosmological
spacetimes \cite{veg1,ven},
cosmic string spacetime \cite{san3}, gravitational
wave backgrounds \cite{san4},
supergravity backgrounds
(which are necessary for fermionic strings) \cite{san5},
and near spacetime singularities \cite{san6}. Physical
phenomena like the Hawking-Unruh
effect in string theory \cite{san1,san7}, horizon
string stretching \cite{san1,san7}, particle
transmutation \cite{san2,san8}, string scattering \cite{san2,san3}, mass
spectrum and critical
dimension \cite{veg1,san2,san3}, string
instability [1, 5-8, 12], multi-string
solutions [6-8]
have been found. Is also useful to consider simple tractable spacetimes
of physical interest, and the restriction to lower dimensions.
Although two dimensional models
have many attractive tractable aspects and can be used to test and get
insights on particular features, D=2 is not for string
theory neither for gravity a
physically appealing dimension \cite{san9}.
In contrast, D=2+1 posses all the physical ingredients
of string theory and gravity in
higher dimensions [6-8, 20-26].

In this paper we investigate the string dynamics in the 2+1 black hole anti
de Sitter (BH-ADS) spacetime recently found by Ba\~{n}ados et. al. \cite{ban1}.
This spacetime background has arised much interest recently
[21-25]. It
describes a two-parameter family (mass $M$ and angular momentum $J$) of black
holes in 2+1 dimensional general relativity with metric:
\begin{equation}
ds^2=(M-\frac{r^2}{l^2})dt^2+(\frac{r^2}{l^2}-M+\frac{J^2}{4r^2})^{-1}dr^2
-Jdtd\phi+r^2 d\phi^2.
\end{equation}
It has two horizons
$r_\pm=\sqrt{\frac{Ml^2}{2}\pm\frac{l}{2}\sqrt{M^2 l^2-J^2}}$ and a
static limit $r_{\mbox{erg}}=\sqrt{M}l,$ defining
an ergosphere, as for ordinary Kerr
black holes. The spacetime is not asymptotically flat however, it approaches
anti de Sitter spacetime asymptotically with cosmological constant
$\Lambda=-1/l^2.$ The curvature is constant $R_{\mu\nu}=-(2/l^2)g_{\mu\nu}$
everywhere, except probably at $r=0,$ where it has at most a delta-function
singularity. Notice the weak nature of the singularity at $r=0$ in 2+1
dimensions as compared with the power law divergence of curvature scalars in
$D>3$ (We will not discuss here the geometry near $r=0.$ For a discusion,
see Refs.\cite{ban2,delta}). The
spacetime, eq. (1.1), is also a solution of the low
energy effective action of string theory with zero dilaton field $\Phi=0,$
anti-symmetric tensor field $H_{\mu\nu\rho}=(2/l^2)\epsilon_{\mu\nu\rho}\;\;
(\mbox{i.e.}\;B_{\phi t}=r^2/l^2)$ and $k=l^2$ \cite{hor1}. Moreover, it
yields an
exact solution of string theory in 2+1 dimensions, obtained by gauging the
WZWN sigma model of the group $SL(2,R)\times R$ at level
$k$ \cite{hor1,kal} (for
non-compact groups, $k$ does not need to be an integer, so the central charge
$c=3k/(k-2)=26$ when k=52/23). This solution is the black string background
\cite{hor2}:
\begin{eqnarray}
&d\tilde{s}^2=-(1-\frac{{\cal M}}{\tilde{r}})d\tilde{t}^2+(1-\frac{{\cal Q}^2}
{{{\cal M}}\tilde{r}})d\tilde{x}^2+(1-\frac{{\cal M}}{\tilde{r}})^{-1}
(1-\frac{{\cal Q}^2}{{{\cal M}}\tilde{r}})^{-1}\frac{l^2 d\tilde{r}^2}
{4\tilde{r}^2},&\nonumber
\end{eqnarray}
\begin{eqnarray}
&\tilde{B}_{\tilde{x}\tilde{t}}=\frac{{\cal Q}}{r},\;\;\;
\tilde{\Phi}=-\frac{1}{2}
\log\tilde{r}l,&
\end{eqnarray}
which is related by duality \cite{hor1,bus} to
the 2+1 BH-ADS spacetime, eq. (1.1). It
has two horizons $\tilde{r}_{\pm}=r_\pm,$ the same as the metric, eq. (1.1),
while the static limit is $\tilde{r}_{\mbox{erg}}=J/(2\sqrt{M}).$
Throughout the paper we use sign-conventions of Misner-Thorne-Wheeler
\cite{mis}
and units where $G=1,$ $c=1$ and the string tension $(2\pi\alpha')^{-1}=1.$
\vskip 6pt
\hspace*{-6mm}
We first investigate the string propagation in these backgrounds by considering
the perturbation series around the exact center of mass of the string:
\begin{equation}
x^\mu(\tau,\sigma)=q^\mu(\tau)+\eta^\mu(\tau,\sigma)+\xi^\mu(\tau,\sigma)+...
\end{equation}
The original method of Refs.\cite{veg1,san1} can be conveniently formulated in
covariant form. Is also useful to introduce $D-1$ normal vectors
$n^\mu_R\;\;(R=1,..,D-1),$ (which can be chosen to be covariantly constant
by gauge fixing), and consider comoving perturbations $\delta x_R,$ i.e. those
seen by an observer travelling with the center of mass, thus $\eta^\mu=
\delta x^Rn^\mu_R.$ After Fourier transforming $\delta x^R(\tau,\sigma)=
\sum_n C^R_n(\tau)e^{-in\sigma},$ the first order perturbations satisfy the
matrix Schr\"{o}dinger-type equation in $\tau$:
\begin{equation}
\ddot{C}_{nR}+(n^2\delta_{RS}-R_{\mu\rho\sigma\nu}n^\mu_R n^\nu_S
\dot{q}^\rho\dot{q}^\sigma)C^S_n=0.
\end{equation}
Second order perturbations $\xi^\mu$ and constraints are similarly covariantly
treated, $\xi^\mu$ also satisfying Schr\"{o}dinger-type equations
with source terms, see eqs. (3.15)-(3.20).

For our purposes here it is enough to consider the non-rotating $(J=0)$ 2+1
BH-ADS background and a radially infalling string. We solve completely the
c.m. motion $q^\mu(\tau)$ and the first and second order perturbations
$\eta^\mu(\tau,\sigma)$ and $\xi^\mu(\tau,\sigma)$ in this background.
Eq. (1.4) becomes:
\begin{equation}
\ddot{C}_{nR}+(n^2+\frac{m^2}{l^2})C_{nR}=0;\;\;\;R=\perp,\parallel
\end{equation}
The first order perturbations are independent of the black hole mass, only
the ADS part emerges. All oscillation frequencies $\omega_n=
\sqrt{n^2+m^2/l^2}$ are real, there are no unstable modes in this case, the
perturbations:
\begin{equation}
\delta x_R(\tau,\sigma)=\sum_n
[ A_{nR}e^{-i(n\sigma+\omega_n\tau)}+\tilde{A}_{nR}
e^{-i(n\sigma-\omega_n\tau)}]
\end{equation}
are completely finite and regular. This is also true for the second order
perturbations, which are bounded everywhere even for $r\rightarrow 0\;(\tau
\rightarrow 0).$ We also compute the conformal generators $L_n,$ eq. (3.61),
and the string mass:
\begin{equation}
m^2=2\sum_n (2n^2+\frac{m^2}{l^2})[A_{n\parallel}
\tilde{A}_{-n\parallel}+
A_{n\perp}\tilde{A}_{-n\perp}].
\end{equation}
The mass formula is modified (by the term $m^2/l^2$) with respect to the
usual flat spacetime expression. This is due to the asymptotic (here ADS)
behaviour of the spacetime. In ordinary $D>3$ black hole spacetimes (without
cosmological constant), which are asymptotically flat, the mass spectrum
is the same as in flat space \cite{san2}. The
quantum string dynamics and mass spectrum
for the 2+1 BH-ADS spacetime are to be discussed elsewhere.

We compare with the string perturbations in the ordinary ($D>3$) black hole
anti de Sitter spacetime. In this case eqs. (1.4) become:
\begin{equation}
\ddot{C}_{nS\perp}+(n^2+m^2 H^2+\frac{Mm^2}{r^3})
C_{nS\perp}=0,\;\;\;S=1,2
\end{equation}
\begin{equation}
\ddot{C}_{n\parallel}+(n^2+m^2 H^2-\frac{2Mm^2}{r^3})
C_{n\parallel}=0.
\end{equation}
The transverse $\perp$-perturbations are oscillating with real frequencies
and are bounded even for $r\rightarrow 0.$ For
longitudinal $\parallel$-perturbations, however, imaginary
frequencies arise and instabilities develop.
The $(\mid n\mid=1)$-instability sets in at:
\begin{equation}
r_{\mbox{inst.}}=(\frac{2Mm^2}{1+m^2H^2})^{1/3}
\end{equation}
Lower modes become unstable even outside the horizon, while higher modes
develop instabilities at smaller $r$ and eventually only for $r\approx 0.$
For $r\rightarrow 0$ (which implies $\tau\rightarrow\tau_0$) we find $r(\tau)
\approx (3m\sqrt{M/2})^{2/3}(\tau_0-\tau)^{2/3}$ and:
\begin{equation}
\ddot{C}_{nS\perp}+\frac{2}{9(\tau-\tau_0)^2}C_{nS\perp}=0,\;\;\;S=1,2
\end{equation}
\begin{equation}
\ddot{C}_{n\parallel}-\frac{4}{9(\tau-\tau_0)^2}C_{n\parallel}=0.
\end{equation}
For $\tau\rightarrow\tau_0$ the $\parallel$-perturbations blow up while
the string
ends trapped into the $r=0$ singularity. We see the important difference
between the string evolution in the 2+1 BH-ADS background and the ordinary
3+1 (or higher dimensional) black hole anti de Sitter spacetime.

We also compare with the string propagation in the 2+1 black string background,
eq. (1.2) (with $J=0$). In this case, eqs. (1.4) become:
\begin{equation}
\ddot{C}_{n\perp}+n^2 C_{n\perp}=0,
\end{equation}
\begin{equation}
\ddot{C}_{n\parallel}+(n^2-\frac{2m^2M}{lr})C_{n\parallel}=0.
\end{equation}
The $\perp$-modes are stable, while $C_{n\parallel}$ develop instabilities. For
$r\rightarrow 0$ (which implies $\tau\rightarrow\tau_0$) we find $r(\tau)
\approx\frac{m^2M}{l}(\tau_0-\tau)^2$ and:
\begin{equation}
\ddot{C}_{n\parallel}-\frac{2}{(\tau_0-\tau)^2}C_{n\parallel}=0,
\end{equation}
with similar conclusions as for the ordinary 3+1 (or higher dimensional)
black hole anti
de Sitter spacetime.
\vskip 6pt
\hspace*{-6mm}In order to extract more information about the string evolution
in these backgrounds, in particular exact properties, we consider the
circular string ansatz:
\begin{equation}
t=t(\tau),\;\;\;r=r(\tau),\;\;\;\phi=\sigma+f(\tau),
\end{equation}
in the equatorial plane $(\theta=\pi/2)$ of the stationary axially symmetric
backgrounds:
\begin{equation}
ds^2=g_{tt}(r)dt^2+g_{rr}(r)dr^2+2g_{t\phi}(r)dtd\phi+g_{\phi\phi}(r)d\phi^2.
\end{equation}
This includes
all the cases of interest here: The 2+1 BH-ADS spacetime, the black
string, as well as the equatorial plane of
ordinary Einstein black holes. The string dynamics is
then reduced to a system of second order ordinary differential equations and
constraints, also described as a Hamiltonian system:
\begin{equation}
\dot{r}^2+V(r)=0;\;\;\;\;\;V(r)=g^{rr}(g_{\phi\phi}+E^2g^{tt}),
\end{equation}
\begin{equation}
\dot{t}=-Eg^{tt},\;\;\;\;\;\dot{f}=-Eg^{t\phi};\;\;\;\;E=-P_t=\mbox{const.},
\end{equation}
which in all backgrounds considered
here are solved in terms of either elementary
or elliptic functions. The dynamics of the circular strings takes place at
the $r$-axis in the $(r,V(r))$ diagram and from the properties of the potential
$V(r)$ (minima, zeroes, asymptotic behaviour for large $r$ and the value
$V(0)$), general knowledge about the string motion can be
obtained. On the other hand, the line element of the circular string turns
out to be:
\begin{equation}
ds^2=g_{\phi\phi}(d\sigma^2-d\tau^2),\;\;\;\mbox{i.e.}\;\;S(\tau)=\sqrt
{g_{\phi\phi}(r(\tau))},
\end{equation}
$S(\tau)$ being the invariant string size. For all the static black hole ADS
spacetimes ( 2+1 and higher dimensional) $\;S(\tau)=r(\tau),$ while for the
black string background $S(\tau)=r(\tau)^{-1}$!, reflecting the dual
properties of the background on the circular test string.

For the rotating 2+1 BH-ADS spacetime:
\begin{equation}
V(r)=r^2(\frac{r^2}{l^2}-M)+\frac{J^2}{4}-E^2,
\end{equation}
(see Fig.1). $V(r)$ has a global minimum $V_{\mbox{min}}<0$ between the two
horizons $r_+,r_-$ (for $Ml^2\geq J^2,$ otherwise there are no horizons). The
vanishing of $V(r)$ at $r=r_{01,2}$ (see eq. (4.18)) determines three
different types of solutions: (i) For $J^2>4E^2,$ there are two
positive zeroes $r_{01}<r_{02},$ the string never comes outside the static
limit, never falls into $r=0$ neither (there is a barrier between $r=r_{01}$
and $r=0$). The mathematical solution oscillates between $r_{01}$ and
$r_{02}$ with $0<r_{01}<r_-<r_+<r_{02}<r_{\mbox{erg}}.$ It may be
interpreted as a string travelling between the different universes described
by the maximal analytic extension of the manifold. (ii) For $J^2<4E^2,$ there
is only one positive zero $r_0$ outside the static limit and there is no
barrier preventing the string from collapsing into $r=0$. The string starts
at $\tau=0$ with maximal size $S^{(ii)}_{\mbox{max}}$ outside the static
limit, it then contracts through the ergosphere and the two horizons
and eventually collapses into a point $r=0.$ For $J\neq 0,$ it may be
still possible to continue this solution into another universe like in
the case (i). (iii) $J^2=
4E^2$ is the limiting case where the maximal string size equals the static
limit: $S^{(iii)}_{\mbox{max}}=l\sqrt{M}.$ In this case $V(0)=0,$ thus the
string contracts through the two horizons and eventually collapses into a
point $r=0.$

The exact general solution in the three cases (i)-(iii) is given by:
\begin{equation}
r(\tau)=\mid r_m-\frac{1}{c_1\wp(\tau-\tau_0)+c_2}\mid,
\end{equation}
where:
\begin{equation}
r_m=S_{\mbox{max}}=\sqrt{\frac{Ml^2}{2}}\sqrt{1+\sqrt{1-\frac{4V(0)}{Ml^2}}},
\;\;\;V(0)=\frac{J^2}{4}-E^2.
\end{equation}
$c_1,c_2$ are constants in terms of $(l,M,r_m),$ given by eqs. (4.21), and
$\wp$ is the Weierstrass elliptic $\wp$-function with invariants $(g_2,g_3),$
discriminant $\Delta$ and roots $(e_1,e_2,e_3),$ given by eqs. (4.22)-(4.25).
The three cases (i)-(iii) correspond to the cases $\Delta>0,$ $\Delta<0$ and
$\Delta=0,$ respectively. Notice that
$S^{(ii)}_{\mbox{max}}>S^{(iii)}_{\mbox{max}}=l\sqrt{M}>S^{(i)}_{\mbox{max}}.$
In the case (i), $r(\tau)$ can be written in terms of the Jacobian elliptic
function $\mbox{sn}[\tau^*,k],\;\tau^*=\sqrt{e_1-e_3}\;\tau,\;k=\sqrt{(e_2-
e_3)/(e_1-e_3)}.$ It
follows that the solution (i) oscillates between the two
zeroes $r_{01}$ and $r_{02}$ of $V(r),$ with
period $2\omega,$ where $\omega$ is
the real semi-period of the Weierstrass
function: $\omega=K(k)/\sqrt{e_1-e_3}\;$ in
terms of the complete elliptic integral of the first kind $K(k).$ We have:
\begin{equation}
r(0)=r_m,\;\;\;r(\omega)=\sqrt{Ml^2-r_m^2},\;\;\;r(2\omega)=r_m,...
\end{equation}
In the case (ii) ($\Delta<0$) two roots ($e_1,e_3$) become complex, the string
collapses into a point $r=0$ and we have:
\begin{equation}
r(0)=r_m,\;\;\;r(\frac{\omega_2}{2})=0,\;\;\;r(\omega_2)=r_m,...
\end{equation}
where $\omega_2$ is the real semi-period of the Weierstrass function for this
case. In the case (iii) ($\Delta=0$) the elliptic functions reduce to
hyperbolic functions:
\begin{equation}
r(\tau)=\frac{\sqrt{M}l}{\cosh(\sqrt{M}\tau)},
\end{equation}
so that:
\begin{equation}
r(-\infty)=0,\;\;\;r(0)=r_m=\sqrt{M}l,\;\;\;r(+\infty)=0.
\end{equation}
Here, the string starts as a point, grows until $r=r_m$
(at $\tau=0$), and then it contracts until it collapses again ($r=0$) at
$\tau=+\infty.$ In this case the string makes only
one oscillation between $r=0$ and $r=r_m.$

Notice that for the static background ($J=0$), the only allowed motion is (ii),
i.e. $r_m>r_{\mbox{hor}}=\sqrt{M}l$ (there is no ergosphere and only one
horizon in this case), with:
\begin{equation}
r_m=\sqrt{\frac{Ml^2}{2}}\sqrt{1+\sqrt{1+\frac{4E^2}{Ml^2}}}.
\end{equation}
For $J=0,$ the string collapses into $r=0$ and stops there. The Penrose
diagram of the 2+1 BH-ADS spacetime for $J=0$ is very similar to the
Penrose diagram of the ordinary ($D>3$) Schwarzschild  spacetime, so the
string motion outwards from $r=0$ is unphysical because of the causal
structure. The coordinate time $t(\tau)$ is expressed in terms of the
incomplete elliptic integral of the third kind $\Pi,$ eq. (4.40). The
string has its maximal size $r_m$ at $\tau=0,$ passes the horizon at
$\tau=\tau_{\mbox{hor}}$ (expressed in terms of the incomplete elliptic
integral of the first kind, eq. (4.41)) and falls into $r=0$ for $\tau=
\omega_2/2,$ $\omega_2$ being the real semi-period of the Weierstrass
function, eq. (4.36). That is, we have:
\begin{eqnarray}
&r(0)=r_m,\;\;\;r(\tau_{\mbox{hor}})=\sqrt{M}l,\;\;\;r(\omega_2/2)
=0,&\nonumber\\
&t(0)=0,\;\;\;t(\tau_{\mbox{hor}})=\infty,&
\end{eqnarray}
and $t(\omega_2/2)$ is expressed in terms of the Jacobian zeta function
$Z,$ eq. (4.42).
\vskip 6pt
\hspace*{-6mm}We also study the circular strings in the ordinary $D>3$
spacetimes. In the 3+1 Kerr anti de Sitter (or Kerr de Sitter) spacetime,
the potential $V(r)$ is given by eq. (4.46), covering seven powers in $r.$
The general circular string solution involves higher genus elliptic functions
and it is not necessary to go into details here. We will compare with the
non-rotating cases, only.

It is instructive to recall \cite{vil4} the circular
string in Minkowski spacetime (MIN), for
which $V(r)=r^2-E^2$ (Fig.2a), the string
oscillates between its maximal size $r_m=E,$ and $r=0$ with the solution
$r(\tau)=r_m\mid\cos\tau\mid.$

In the Schwarzschild black hole (S)
$V(r)=r^2-2Mr-E^2$ (Fig.2b), the solution is remarkably simple: $r(\tau)=M+
\sqrt{M^2+E^2}\cos\tau.$ The mathematical solution oscillates between
$r_m=M+\sqrt{M^2+E^2}$ and $M-\sqrt{M^2+E^2}<0,$ but because of the causal
structure and the curvature singularity the motion can not be continued after
the string has collapsed into $r=0.$

For anti de Sitter spacetime (ADS), we
find: $V(r)=r^2(1+H^2r^2)-E^2$ (Fig.2c). The
string oscillates between $r_m=\frac{1}{\sqrt{2}H}\sqrt{-1+\sqrt{1+4H^2E^2}}$
and $r=0$ with the solution:
\begin{equation}
r(\tau)=r_m\mid\mbox{cn}[(1+4H^2E^2)^{1/4},k]\mid,
\end{equation}
which is periodic with period $2\omega:$
\begin{equation}
\omega=\frac{K(k)}{(1+4H^2E^2)^{1/4}},\;\;\;\;k=\sqrt{\frac{\sqrt{1+4H^2E^2}-1}
{2\sqrt{1+4H^2E^2}}}.
\end{equation}

For Schwarzschild anti de Sitter spacetime (S-ADS), we find $V(r)=
r^2(1+H^2r^2)-2Mr-E^2$ (Fig.2d) and:
\begin{equation}
r(\tau)=r_m-\frac{1}{d_1\wp(\tau)+d_2},\;\;\;\;r(0)=r_m
\end{equation}
$d_1,d_2$ are constants given in terms of $(M,H,r_m)$ by eqs. (4.60) and
(4.62) ($r_m$ is the root of the equation $V(r)=0$,  which has
in this case exactly one positive solution). The invariants, the
discriminant and the roots are determined by
eqs. (4.63)-(4.64). The string starts with $r=r_m$
at $\tau=0,$ it then contracts and eventually collapses into the $r=0$
singularity. The existence of elliptic function solutions for the
string motion is characteristic of the presence of a cosmological constant.
For $\Lambda=0=-3H^2$ the circular string motion reduces to simple
trigonometric functions. From Fig.2 and our analysis we see that the circular
string motion is
qualitatively very similar in all these backgrounds (MIN, S, ADS,
S-ADS): the string has a maximal {\it bounded} size
and then it contracts towards
$r=0.$ There are however physical and quantitative differences: in Minkowski
and pure anti de Sitter spacetimes, the string truly oscillates between
$r_m$ and $r=0,$ while in the black hole cases (S, S-ADS), there is only one
half oscillation, the string motion stops at $r=0.$ This also holds for the
2+1 BH-ADS spacetime with $J=0$ (Fig.1b). Notice also that in all these
cases, $V(0)=-E^2<0$ and $V(r)\sim r^\alpha;\;(\alpha=2,4)$ for $r>>E.$

The similarity can be pushed one step further by considering small
perturbations
around the circular strings. We find:
\begin{equation}
\ddot{C}_{n}+(n^2+\frac{r}{2}\frac{da(r)}{dr}+\frac{r^2}{2}\frac{d^2a(r)}
{dr^2}-\frac{2E^2}{r^2})C_n=0,
\end{equation}
determining the Fourier components of the comoving perturbations. For the
spacetimes of interest here, $a(r)=1-2M/r+H^2r^2$ (MIN, S, ADS, S-ADS), or
$a(r)=r^2/l^2-1$ (2+1 BH-ADS), the comoving perturbations are regular except
near $r=0,$ where we find (for all cases) $r(\tau)\approx-E(\tau-\tau_0)$ and:
\begin{equation}
\ddot{C}_{n}-\frac{2}{(\tau-\tau_0)^2}C_n=0.
\end{equation}
It follows that not only the unperturbed circular strings, but
also the comoving
perturbations around them behave in a similar way in all these non-rotating
backgrounds (2+1 and higher dimensional). This should be contrasted with the
string perturbations around the center of mass, which behave differently in
these backgrounds. It must be noticed that for rotating ($J\neq 0$) spacetimes,
the circular string behaviour is qualitatively different from the non-rotating
($J=0$) spacetimes. For large $J,$ both in the 2+1 BH-ADS as well as in the
3+1 ordinary Kerr-ADS spacetimes, non-collapsing circular string solutions
exist. The potential $V(r)\rightarrow +\infty$ for $r\rightarrow 0$ and no
collapse into $r=0$ is possible.

The dynamics of circular strings in curved spacetimes is determined by the
string tension, which tends to contract the string, and by the local gravity
(which may be attractive or repulsive). In all the previous backgrounds, the
local gravity is attractive (i.e. $da(r)/dr>0$), and it acts together with
the string tension in the sense of contraction. But in spacetimes with regions
in which repulsion (i.e. $da(r)/dr<0$) dominates, the strings can expand with
unbounded radius (unstable strings \cite{mic,veg4}). It
may also happen that the
string tension and the local gravity be of the same order, i.e. the
two opposite
effects can balance, and the string is stationary. De Sitter spacetime provides
an example in which all such type of solutions exist \cite{mic,veg4}. In
de Sitter
spacetime, $V(r)$ is unbounded from below for $r\rightarrow\infty$ ($V(r)\sim
-r^4$) and unbounded expanding circular strings are present. In addition, an
interesting new feature appears in the presence of a
positive cosmological constant: the existence of multi-string
solutions [6-8]. The
world-sheet time $\tau$ turns out to be a multi
(finite or infinite) valued function of the physical time. That is, one single
world-sheet where $-\infty\leq\tau\leq+\infty,$ can describe many (even
infinitely many \cite{veg4})
different and independent strings (in flat spacetime,
one single world-sheet describes only one string). In the S, ADS and S-ADS
spacetimes, the multi-string feature is absent.

We also study here the
circular strings in Schwarzschild de Sitter spacetime, where regions with
$da(r)/dr>0$ and $da(r)/dr<0$ exist. The potential in this case is:
$V(r)=-H^2r^4+r^2-2Mr-E^2$ with $V(0)=V(r_+)=V(r_-)=-E^2,$ where the
horizons $r_\pm$ are given by eqs. (5.6)-(5.8), and $V(r)
\sim -r^4$ for large $r,$ see Fig.3. It has a local minimum between $r=0$ and
$r_-,$ and a local maximum at $r=r_o$ given by eq. (5.12), $r_-<r_o<r_+.$ The
motion is very complicated here, but again, it can be exactly determined
in terms of elliptic functions, for which we analyze here only the degenerate
case. We find two different types of solutions $r_\pm(\tau)$ given by eqs.
(5.15)-(5.19) with the properties:
\begin{eqnarray}
&r_+(-\infty)=r_o,\;\;\;\;r_+(0)=0,&\nonumber\\
&r_-(-\infty)=r_o,\;\;\;\;r_-(0)=\infty,\;\;\;\;r_-(\infty)=r_o.&
\end{eqnarray}
$r_+(\tau)$ describes one contracting string starting with maximal size $r_o$
at $\tau=-\infty,$ passing the inner horizon $r_-$ and falling into the
$r=0$ singularity at $\tau=0.$ The solution $r_-(\tau)$ describes two
different and independent strings: The string I starts with minimal size $r_o$
at $\tau=-\infty$ and grows until infinite size at $\tau=0$ (unstable string).
String II starts with infinite size at $\tau=0$ and contracts till minimal
size $r_o$ at $\tau=+\infty.$ They never collapse into the $r=0$ singularity.
The $r_-(\tau)$ solution is very similar to the two-string solution
discussed in Refs.\cite{mic,veg4} in the pure de Sitter case.
\vskip 6pt
\hspace*{-6mm}Finally, we study the circular string in the black string
background. In this case (see Fig.4):
\begin{equation}
\tilde{V}(r)=\frac{J^2}{4r^4}-\frac{M}{r^2}+\frac{1}{l^2}-E^2,
\end{equation}
with $V(r_\pm)=-E^2,\;\;V(\infty)=1/l^2-E^2$ and $V(0)=+\infty$ for $J\neq 0,$
while $V(0)=-\infty$ for $J=0$ (we only consider positive $M$). One effect
of the dual transformation is to change the asymptotic behaviour of $V(r).$ We
see that if $E^2l^2>1,$ then $V(\infty)<0,$ which gives rise to solutions of
unbounded $r.$ This is to be contrasted with the solutions in the BH-ADS
spacetimes in which the ring solutions are always bounded. Another effect of
duality  here, is to change the invariant string size, we find $\tilde{S}(\tau)
=1/r(\tau).$ For $J\neq 0,$ all solutions are bounded (finite $\tilde{S}$),
while for $J=0,$ unbounded (infinite $\tilde{S}$) exist as well. For $J\neq 0,$
the general solution can be expressed in terms of elliptic functions
(elementary functions for $J=0$), whose description and physical
interpretation are to be described elsewhere.
\vskip 12pt
\hspace*{-6mm}This paper is organized as follows: In Section 2, we review the
2+1 BH-ADS and black string backgrounds. In Section 3 we describe and solve the
string perturbations around the string center of mass in these backgrounds and
in the ordinary black hole ADS spacetime. In Section 4 we solve the exact
circular string motion in all these backgrounds and compare between them, and
in Section 5 we discuss the circular string motion in the black hole de Sitter
case. A summary of our results and conclusions is presented in Tables I
and II.
\section{Review of the 2+1 Dimensional Black Hole}
\setcounter{equation}{0}
In this section we give a short introduction to the black hole anti de Sitter
(BH-ADS) solution of $2+1$ dimensional Einstein theory, recently found by
Ba\~{n}ados et. al. \cite{ban1}. There are now several ways to obtain this
solution [20-22]. A simple way is to take as the starting point
a line element in the form:
\begin{equation}
ds^2=-a(r)dt^2+\frac{dr^2}{a(r)}+r^2 d\phi^2,
\end{equation}
where $a(r)$ is an arbitrary function of $r$. The non-vanishing components of
the Einstein tensor, $G_{\mu\nu}=R_{\mu\nu}-\frac{R}{2}g_{\mu\nu}$, take
the form:
\begin{equation}
G_{rr}=\frac{a_{,r}}{2ar},\;\;\;G_{tt}=-\frac{aa_{,r}}
{2r},\;\;\;G_{\phi\phi}=
\frac{r^2}{2}a_{,rr}.
\end{equation}
The only vacuum solution to the
Einstein equations is $a=\mbox{const}$, corresponding to
flat spacetime, as is well-known in 3 dimensions. However, if we introduce a
cosmological constant:
\begin{equation}
T^\mu\;_\nu=\mbox{diag}(1,1,1)\Lambda,
\end{equation}
the solution to the Einstein equations becomes non-trivial:
\begin{equation}
a(r)=c-\Lambda r^2,
\end{equation}
where $c$ is an arbitrary constant. Usually $c$ is scaled to $1$, and then
a positive $\Lambda$ represents de Sitter space, while a negative $\Lambda$
represents anti de Sitter space.
On the other hand, since the constant $c$ is completely arbitrary, we may
as well take a negative $c$ and then we also find solutions in the form:
\begin{equation}
ds^2=(1-\frac{r^2}{l^2})dt^2+(\frac{r^2}{l^2}-1)^{-1}dr^2+r^2 d\phi^2,
\end{equation}
where $l$ is a constant. This is in fact the simplest example of the $2+1$
BH-ADS solutions of Ref.\cite{ban1}. This particular solution, where $\phi$
is identified with $\phi+2\pi$, is a black hole spacetime with mass $M=1$
and angular momentum $J=0$. There is a horizon at $r=l$ and asymptotically
it approaches anti de Sitter space with $\Lambda=-1/l^2$. A two-parameter
family (mass $M$ and angular momentum $J$) of black holes is obtained by
periodically identifying a linear combination of $t$ and $\phi$. This leads
to the solution (the details can be found in Refs.\cite{ban1,hor1}):
\begin{equation}
ds^2=(M-\frac{r^2}{l^2})dt^2+(\frac{r^2}{l^2}-M+\frac{J^2}{4r^2})^{-1}dr^2
-Jdtd\phi+r^2 d\phi^2,
\end{equation}
with two horizons (provided $Ml^2>J^2$):
\begin{equation}
r_\pm=\sqrt{\frac{Ml^2}{2}\pm\frac{l}{2}\sqrt{M^2 l^2-J^2}}
\end{equation}
and a static limit:
\begin{equation}
r_{\mbox{erg}}=\sqrt{M}l,
\end{equation}
defining an ergosphere, as for ordinary Kerr black holes.
The Riemann tensor,
corresponding to the line element (2.6), is given by:
\begin{equation}
R_{\mu\rho\sigma\nu}=-\frac{1}{l^2}(g_{\mu\sigma}g_{\nu\rho}-g_{\mu\nu}
g_{\sigma\rho}),
\end{equation}
so that the curvature is constant:
\begin{equation}
R_{\mu\nu}=-\frac{2}{l^2}g_{\mu\nu}.
\end{equation}
The geometry of the solution (2.6), near $r=0$ in particular, is discussed
in detail in Refs.[23, 25].
\vskip 6pt
\hspace*{-6mm}We
close this section with a few remarks on the relevance of this solution
in string theory \cite{hor1,kal}. To lowest order in an expansion in
$\alpha'$, the string action is \cite{gib}:
\begin{equation}
S=\int d^3 x\sqrt{-g}e^{-2\Phi}[\frac{4}{k}+R+4(\nabla\Phi)^2-\frac{1}{12}
H_{\mu\nu\rho}H^{\mu\nu\rho}],
\end{equation}
where $\Phi$ is the dilaton field and $H_{\mu\nu\rho}$ is the field strength
of the Kalb-Ramond field $B_{\mu\nu},\;(H_{\mu\nu\rho}=\partial_{[\rho}
B_{\mu\nu]})$. It is easy to show that the metric
(2.6) is a solution to the equations of motion corresponding to the action
(2.11), when supplemented by \cite{hor1}:
\begin{equation}
B_{\phi t}=\frac{r^2}{l^2},\;\;\;\Phi=0,\;\;\;k=l^2.
\end{equation}
Following Horowitz and Welch \cite{hor1} the connection to the WZWN
$\sigma$-model approach is most easily established by dualizing the solution
(2.6),(2.12) on the cyclic coordinate $\phi$. According to the well-known
procedure \cite{bus}, the dual solution to (2.6),(2.12) is then given
by \cite{hor1}:
\begin{eqnarray}
&d\tilde{s}^2=(M-\frac{J^2}{4r^2})dt^2+(\frac{r^2}{l^2}-M+\frac{J^2}{4r^2})^{-1}
dr^2+\frac{2}{l}dtd\phi+\frac{d\phi^2}{r^2},&\nonumber
\end{eqnarray}
\begin{eqnarray}
&\tilde{B}_{\phi t}=\frac{-J}{2r^2},\;\;\;\tilde{\Phi}=-\log r,&
\end{eqnarray}
that after diagonalization of the metric becomes:
\begin{eqnarray}
&d\tilde{s}^2=-(1-\frac{{\cal M}}{\tilde{r}})d\tilde{t}^2+(1-\frac{{\cal Q}^2}
{{{\cal M}}\tilde{r}})d\tilde{x}^2+(1-\frac{{\cal M}}{\tilde{r}})^{-1}
(1-\frac{{\cal Q}^2}{{{\cal M}}\tilde{r}})^{-1}\frac{l^2 d\tilde{r}^2}
{4\tilde{r}^2},&\nonumber
\end{eqnarray}
\begin{eqnarray}
&\tilde{B}_{\tilde{x}\tilde{t}}=\frac{{\cal Q}}{r},\;\;\;
\tilde{\Phi}=-\frac{1}{2}
\log\tilde{r}l,&
\end{eqnarray}
where:
\begin{eqnarray}
&t=\frac{l(\tilde{x}-\tilde{t})}{\sqrt{r^2_+-r^2_-}},\;\;\;\phi=\frac{r^2_+
\tilde{t}-
r^2_-\tilde{x}}{\sqrt{r^2_+-r^2_-}},&\nonumber
\end{eqnarray}
\begin{eqnarray}
&{\cal M}=\frac{r^2_+}{l},\;\;\;{\cal Q}=\frac{J}{2},\;\;\;r^2=\tilde{r}l.&
\end{eqnarray}
The metric of (2.14) is exactly the black string solution in 3 dimensions of
Horne and Horowitz \cite{hor2}, obtained by gauging the WZWN $\sigma$-model
of the group $SL(2,R)\times R$.

Notice that the spacetime (2.13) is stationary and axially symmetric, and
that it has the same non-vanishing components of the metric tensor as the
original spacetime (2.6). This will be important when we consider circular
strings in Section 5. Another interesting observation is that the duality
transformation does not change the two horizons (2.7), while the static
limit is changed to $r_{\mbox{erg}}(\mbox{black string})=J/(2\sqrt{M})$.
\section{Perturbations Around the String Center of Mass}
\setcounter{equation}{0}
One of the main purposes of the
present paper is to consider the classical propagation of a
bosonic test string in the $2+1$ BH-ADS spacetime
(the spacetime is taken as a fixed
background and no backreactions of the strings are included). The point
particle geodesics were recently
investigated in Ref.\cite{far}.
The string equations of motion are
highly non-linear coupled partial differential equations, so we will
restrict ourselves by considering two different approaches. In this
section we calculate first and second order string fluctuations around the
string center of mass, following the approach originally developed by
de Vega and S\'{a}nchez \cite{veg1}, and in Section 4 we consider
exact circular strings winding around the black hole.
In both cases analysis and comparison with the ordinary black hole
ADS (and DS) solutions in $3+1$
dimensions
is done.
\subsection{General Formalism}
Let us solve the string equations of motion and constraints by considering
perturbations around the exact
string center of mass solution.
In this subsection, we shortly review the method, and
we demonstrate the simplifications arising at first order in the expansion,
when considering only physical (perpendicular to the geodesic) perturbations.
\vskip 6pt
\hspace*{-6mm}In
an arbitrary curved spacetime of dimension $D$, the string equations of
motion and constraints, in the conformal gauge, take the form:
\begin{equation}
\ddot{x}^\mu-x''^\mu+\Gamma^\mu_{\rho\sigma}(\dot{x}^\rho\dot{x}^\sigma-
x'^\rho x'^\sigma)=0,
\end{equation}
\begin{equation}
g_{\mu\nu}\dot{x}^\mu x'^\nu=g_{\mu\nu}(\dot{x}^\mu\dot{x}^\nu+x'^\mu
x'^\nu)=0,
\end{equation}
for $\mu=0,1,...,(D-1)$ and prime and dot represent derivative with
respect to $\sigma$ and $\tau,$ respectively. Consider
first the equations of motion (3.1). A particular solution is
provided by the string center of mass $q^\mu(\tau)$:
\begin{equation}
\ddot{q}^\mu+\Gamma^\mu_{\rho\sigma}\dot{q}^\rho\dot{q}^\sigma=0.
\end{equation}
Then a perturbative series around this solution is developed:
\begin{equation}
x^\mu(\tau,\sigma)=q^\mu(\tau)+\eta^\mu(\tau,\sigma)+\xi^\mu(\tau,\sigma)+...
\end{equation}
After insertion of eq. (3.4) in eq. (3.1) the equations of motion are to
be solved order by order in the expansion.

To zeroth order we just get eq. (3.3). To first order we find:
\begin{equation}
\ddot{\eta}^\mu+\Gamma^\mu_{\rho\sigma,\lambda}\dot{q}^\rho\dot{q}^\sigma
\eta^\lambda+2\Gamma^\mu_{\rho\sigma}\dot{q}^\rho\dot{\eta}^\sigma-
\eta''^\mu=0.
\end{equation}
The first three terms can be written in covariant form \cite{men}, c.f. the
ordinary geodesic deviation equation:
\begin{equation}
\dot{q}^\lambda\nabla_\lambda(\dot{q}^\delta\nabla_\delta\eta^\mu)-
R^\mu_{\epsilon\delta\lambda}\dot{q}^\epsilon\dot{q}^\delta\eta^\lambda-
\eta''^\mu=0.
\end{equation}
However, we can go one step further. For a massive string, corresponding to the
string center of mass fulfilling:
\begin{equation}
g_{\mu\nu}(q)\dot{q}^\mu\dot{q}^\nu=-m^2,
\end{equation}
there are $D-1$ physical polarizations
of string perturbations around the geodesic $q^\mu(\tau)$. We therefore
introduce $D-1$ normal vectors $n^\mu_R,\;R=1,2,...,(D-1)$:
\begin{equation}
g_{\mu\nu}n^\mu_R\dot{q}^\nu=0,\;\;\;g_{\mu\nu}n^\mu_R n^\nu_S=\delta_{RS}
\end{equation}
and consider only first order perturbations in the form:
\begin{equation}
\eta^\mu=\delta x^R n^\mu_R,
\end{equation}
where $\delta x^R$ are the comoving perturbations, i.e. the perturbations
as seen by an observer travelling with the center of mass of the string. The
normal vectors are not uniquely defined by eqs. (3.8). In fact, there
is a gauge invariance originating from the freedom to make local
rotations of the $(D-1)$-bein spanned by the normal vectors. For our purposes
it is convenient to fix the gauge taking
the normal vectors to be covariantly constant:
\begin{equation}
\dot{q}^\mu\nabla_\mu n^\nu_R=0.
\end{equation}
This is achieved by choosing the basis $(q^\mu, n^\mu_R)$ obeying conditions
(3.8) at a given point, and defining it along the geodesic by means of
parallel transport. Another useful formula is the completeness relation
that takes the form:
\begin{equation}
g^{\mu\nu}=-\frac{1}{m^2}\dot{q}^\mu\dot{q}^\nu+n^{\mu R}n^\nu_R.
\end{equation}
Using eqs. (3.7)-(3.10) in eq. (3.6) we find after multiplication by
$g_{\mu\nu}n^\nu_S$ the spacetime invariant formula:
\begin{equation}
(\partial^2_\tau-\partial^2_\sigma)\delta x_R-R_{\mu\rho\sigma\nu}
n^\mu_R n^\nu_S
\dot{q}^\rho\dot{q}^\sigma\delta x^S=0.
\end{equation}
Since the last term depends on $\sigma$ only through $\delta x^S$ it is
convenient to make a Fourier expansion:
\begin{equation}
\delta x_R(\tau,\sigma)=\sum_n C_{nR}(\tau)e^{-in\sigma}
\end{equation}
Then, eq. (3.12) finally reduces to:
\begin{equation}
\ddot{C}_{nR}+(n^2\delta_{RS}-R_{\mu\rho\sigma\nu}n^\mu_R n^\nu_S
\dot{q}^\rho\dot{q}^\sigma)C^S_n=0,
\end{equation}
which constitutes a matrix Schr\"{o}dinger equation with $\tau$ playing the
role of the spatial coordinate.
\vskip 6pt
\hspace*{-6mm}For the
second order perturbations the picture is a little more complicated.
Since they couple to the first order perturbations we consider the full set
of perturbations $\xi^\mu$ \cite{veg1,men}:
\begin{equation}
\dot{q}^\lambda\nabla_\lambda(\dot{q}^\delta\nabla_\delta\xi^\mu)-
R^\mu_{\epsilon\delta\lambda}\dot{q}^\epsilon\dot{q}^\delta\xi^\lambda-
\xi''^\mu=U^\mu,
\end{equation}
where the source $U^\mu$ is bilinear in the first order perturbations, and
explicitly given by:
\begin{equation}
U^\mu=-\Gamma^\mu_{\rho\sigma}(\dot{\eta}^\rho\dot{\eta}^\sigma-
\eta'^\rho\eta'^\sigma)-2\Gamma^\mu_{\rho\sigma,\lambda}\dot{q}^\rho
\eta^\lambda\dot{\eta}^\sigma-\frac{1}{2}\Gamma^\mu_{\rho\sigma,
\lambda\delta}\dot{q}^\rho\dot{q}^\sigma\eta^\lambda\eta^\delta.
\end{equation}
\vskip 6pt
After solving eqs. (3.14) and (3.15) for the first and second order
perturbations, the constraints (3.2) have to be imposed. In world-sheet
light cone coordinates $(\sigma^\pm=\tau\pm\sigma)$
the constraints take the form:
\begin{equation}
T_{\pm\pm}=g_{\mu\nu}\partial_\pm x^\mu\partial_\pm x^\nu=0,
\end{equation}
where $\partial_\pm=\frac{1}{2}(\partial_\tau\pm\partial_\sigma)$.
The world-sheet energy-momentum tensor $T_{\pm\pm}$ is conserved, as can be
easily verified using eq. (3.1), and therefore can be written:
\begin{equation}
T_{--}=\frac{1}{2\pi}\sum_n\tilde{L}_n e^{-in(\sigma-\tau)},\;\;\;T_{++}=
\frac{1}{2\pi}\sum_n L_n e^{-in(\sigma+\tau)}.
\end{equation}
At the classical level under consideration in this paper, the constraints are
then simply:
\begin{equation}
L_n=\tilde{L}_n=0;\;\;\; n\in Z.
\end{equation}
Up to second order in the expansion around the string center of mass we find:
\begin{eqnarray}
T_{\pm\pm}\hspace*{-2mm}&=&\hspace*{-2mm}-\frac{1}{4}m^2\alpha'^2+g_{\mu\nu}
\dot{q}^\mu\partial_\pm\eta^\nu
+\frac{1}{4}g_{\mu\nu,\rho}\dot{q}^\mu
\dot{q}^\nu\eta^\rho\nonumber\\
\hspace*{-2mm}&+&\hspace*{-2mm}g_{\mu\nu}\dot{q}^\mu\partial_\pm\xi^\nu
+g_{\mu\nu}\partial_\pm\eta^\mu\partial_\pm
\eta^\nu+g_{\mu\nu,\rho}\dot{q}^\mu\eta^\rho\partial_\pm\eta^\nu\nonumber\\
\hspace*{-2mm}&+&\hspace*{-2mm}\frac{1}{4}g_{\mu\nu,\rho}\dot{q}^\mu
\dot{q}^\nu\xi^\rho+\frac{1}{8}g_{\mu\nu,\rho\sigma}\dot{q}^\mu
\dot{q}^\nu\eta^\rho\eta^\sigma
\end{eqnarray}
In the following subsections we apply this formalism to the $2+1$ BH-ADS
and to the black string
solution of Section 2, as well as to ordinary $3+1$ dimensional black hole
ADS solutions.
\subsection{Strings in the 2+1 BH-ADS Background}
We now consider a string in the background of the $2+1$ dimensional BH-ADS
spacetime. For simplicity we take a non-rotating black hole $(J=0)$ and we
consider a radially infalling string. This case is of sufficient complexity
for our purposes; the more general case of a string with angular momentum in
the rotating background is to be considered elsewhere.

Eqs. (3.3) and (3.7) determining the string center of mass lead to:
\begin{equation}
\dot{t}=\frac{-E}{M-\frac{r^2}{l^2}},
\end{equation}
\begin{equation}
\dot{r}^2+m^2(\frac{r^2}{l^2}-M)=E^2,
\end{equation}
where $E$ is an integration constant. These two equations are solved by:
\begin{equation}
t(\tau)=\frac{l}{2\sqrt{M}}\log\mid\frac{1-\frac{E}{m\sqrt{M}}\tan
\frac{m}{l}\tau}{1+\frac{E}{m\sqrt{M}}\tan\frac{m}{l}\tau}\mid
\end{equation}
and:
\begin{equation}
r(\tau)=-\frac{l}{m}\sqrt{Mm^2+E^2}\;\sin\frac{m}{l}\tau.
\end{equation}
Here the boundary conditions were chosen such that $r$ takes its maximal
value at $\tau=-\frac{l}{m}\frac{\pi}{2}$ and the string falls into
$r=0$ for $\tau\rightarrow 0_-$:
\begin{equation}
r_{\mbox{max}}=
r(-\frac{l}{m}\frac{\pi}{2})=\frac{l}{m}\sqrt{Mm^2+E^2},\;\;\;\;
r(0)=0.
\end{equation}
The string center of mass passes the horizon $r_{\mbox{hor}}=\sqrt{M}l$ at:
\begin{equation}
\tau_{\mbox{hor}}=-\frac{l}{m}\arcsin\frac{m\sqrt{M}}{\sqrt{Mm^2+E^2}}
\end{equation}
and we find from eq. (3.23):
\begin{equation}
t(-\frac{l}{m}\frac{\pi}{2})=0,\;\;\;t(\tau_{\mbox{hor}})=\infty,\;\;\;t(0)=0.
\end{equation}
Let us now turn to the string perturbations around the solution (3.23)-(3.24).
The two covariantly constant normal vectors fulfilling eqs. (3.8) and (3.10)
are given by:
\begin{eqnarray}
n^\mu_\perp=(0,\;0,\;\frac{1}{r}),\nonumber
\end{eqnarray}
\begin{eqnarray}
n^\mu_\parallel=(\frac{\dot{r}}{m(M-\frac{r^2}{l^2})},\;
-\frac{E}{m},\;0),
\end{eqnarray}
which define transverse
and longitudinal comoving perturbations through eq. (3.9),
respectively. It is however remarkable that in this case we do not need the
explicit expressions for the normal vectors and for the Riemann tensor to
calculate the first order perturbations (3.14). Using eq. (2.9) and the
normalization equations (3.7) and (3.8) we immediately get:
\begin{equation}
\ddot{C}_{nR}+(n^2+\frac{m^2}{l^2})C_{nR}=0,
\end{equation}
where $R$ takes the values $"\perp"$ and $"\parallel"$. Equations (3.29) are
easily solved and the comoving perturbations (3.13) are given by:
\begin{equation}
\delta x_R(\tau,\sigma)=\sum_n
[ A_{nR}e^{-i(n\sigma+\omega_n\tau)}+\tilde{A}_{nR}
e^{-i(n\sigma-\omega_n\tau)}],
\end{equation}
where:
\begin{equation}
\omega_n=\sqrt{n^2+\frac{m^2}{l^2}},
\end{equation}
\begin{equation}
A_{nR}=\tilde{A}^{\dag}_{-nR}.
\end{equation}
The string perturbations $\eta^\mu$ introduced in eq. (3.4) are:
\begin{equation}
\eta^t=\frac{\dot{r}}{m(M-\frac{r^2}{l^2})}\delta x_\parallel,\;\;\;
\eta^r=-\frac{E}{m}\delta x_\parallel,\;\;\;
\eta^\phi=\frac{1}{r}\delta x_\perp,
\end{equation}
and are plagued by coordinate singularities at $r=r_{\mbox{hor}}$ (for
$\eta^t$) and
at $r=0$ (for $\eta^\phi$). The comoving perturbations
$\delta x_R,$ (3.30), are however
completely finite and regular trigonometric functions. Notice that in the
"pure" de Sitter spacetime \cite{veg1} the perturbations satisfy eq.
(3.30), but with frequency $\omega_n=\sqrt{n^2-m^2/l^2},$ thus unstable modes
(for $\mid n\mid<m/l)$ appear and the perturbations blow up. The presence of
such instabilities is a generic exact feature in the de Sitter spacetime
\cite{sinh,ven}. In
the present 2+1 BH-ADS background, all frequencies $\omega_n$
are real and instabilities do not occur.

Notice also that the comoving perturbations (3.30) are
independent of the black hole mass
$M$ (and of $E$). In fact, eq. (3.29) is the "pure" anti de Sitter result,
where the perturbations are independent of the polarization as they of course
should
be in an isotropic spacetime. This suggests that we have to calculate at least
the second order perturbations to ensure that the effects of the black hole
mass are included in the perturbations. The second order perturbations are
determined by eqs. (3.15)-(3.16). It turns out that the $\xi^\phi$-equation
decouples while the $\xi^t$ and $\xi^r$-equations constitute a set of two
coupled partial second order linear differential equations.

We first consider the $\xi^\phi$-equation. Explicitly it is given by:
\begin{equation}
\ddot{\xi}^\phi-\xi''^\phi+\frac{2}{r}\dot{r}\dot{\xi}^\phi=U^\phi,
\end{equation}
where:
\begin{equation}
U^\phi=-\frac{2}{r}(\dot{\eta}^r\dot{\eta}^\phi-\eta'^r\eta'^\phi)+
2\frac{\dot{r}}{r^2}\eta^r\dot{\eta}^\phi
\end{equation}
The source $U^\phi$ is here written in terms of $\eta^r$ and $\eta^\phi$ and
its explicit expression as a function of $\tau$ and $\sigma$ can be obtained
using eqs. (3.30), (3.33) and (3.24). It is convenient to make the
redefinitions:
\begin{equation}
\Sigma^\phi\equiv r\xi^\phi,\;\;\;\;\tilde{U}^\phi\equiv r U^\phi
\end{equation}
and the Fourier expansions:
\begin{equation}
\Sigma^\phi(\tau,\sigma)=\sum_n\Sigma^\phi_n(\tau)e^{-in\sigma},\;\;\;\;
\tilde{U}^\phi(\tau,\sigma)=\sum_n\tilde{U}^\phi_n(\tau)e^{-in\sigma}.
\end{equation}
Equation (3.34) then reduces to:
\begin{equation}
\ddot{\Sigma}^\phi_n+(n^2+\frac{m^2}{l^2})\Sigma^\phi_n=
\tilde{U}^\phi_n,
\end{equation}
that is solved by:
\begin{eqnarray}
\Sigma^\phi_n(\tau)=B_n e^{-i\omega_n\tau}\hspace*{-2mm}&+&\hspace*{-2mm}
\tilde{B}_n e^{i\omega_n\tau}+\frac{e^{i\omega_n\tau}}{2i\omega_n}
\int^\tau\tilde{U}^\phi_n(\tau')e^{-i\omega_n\tau'}d\tau'\nonumber\\
\hspace*{-2mm}&-&\hspace*{-2mm}\frac{e^{-i\omega_n\tau}}{2i\omega_n}
\int^\tau\tilde{U}^\phi_n(\tau')e^{i\omega_n\tau'}d\tau',
\end{eqnarray}
where $B_n=\tilde{B}^{\dag}_{-n}$ and $\omega_n$ is defined in eq. (3.31).

The perturbations $\xi^t$ and $\xi^r$ are somewhat more complicated to
derive. By redefining $\xi^r$ and $U^r$:
\begin{equation}
\xi^r\equiv(\frac{r^2}{l^2}-M)\xi^*,
\end{equation}
\begin{equation}
U^r\equiv(\frac{r^2}{l^2}-M)U^*,
\end{equation}
we find from eq. (3.15):
\begin{equation}
\left( \begin{array}{c} \ddot{\xi}^t \\  \ddot{\xi}^* \end{array}\right)-
\left( \begin{array}{c} \xi''^t \\ \xi''^* \end{array}\right)+
2{\cal A}\left( \begin{array}{c} \dot{\xi}^t \\ \dot{\xi}^* \end{array}\right)+
{\cal B}\left( \begin{array}{c} \xi^t \\ \xi^* \end{array}\right)=
\left( \begin{array}{c} U^t \\ U^* \end{array}\right),
\end{equation}
where the matrices ${\cal A}$ and ${\cal B}$ are given by:
\begin{eqnarray}
{\cal A}=\left( \begin{array}{cc} a & b \\ b & a \end{array}\right), \;\;\;\;
{\cal B}=\left( \begin{array}{cc} 0 & c \\ 0 & d \end{array}\right);\nonumber
\end{eqnarray}
\begin{eqnarray}
a=\frac{r\dot{r}}{l^2(\frac{r^2}{l^2}-M)},\;\;\;\;b=\frac{Er}{l^2(
\frac{r^2}{l^2}-M)},
\end{eqnarray}
\begin{eqnarray}
c=\frac{2E\dot{r}}{l^2(\frac{r^2}{l^2}-M)},\;\;\;\;d=\frac{2
E^2}{l^2(\frac{r^2}{l^2}-M)}-\frac{m^2}{l^2}.\nonumber
\end{eqnarray}
The first order $\tau$-derivatives in eq. (3.42) are eliminated by the
transformation:
\begin{equation}
\left( \begin{array}{c} \xi^t \\ \xi^* \end{array}\right)\equiv
{\cal G}\left( \begin{array}{c} \Sigma^t \\ \Sigma^*
\end{array}\right);\;\;\;\;
{\cal G}=e^{-\int^\tau{\cal A}(\tau')d\tau'},
\end{equation}
i.e.:
\begin{equation}
{\cal G}=\frac{-1}{m(\frac{r^2}{l^2}-M)}\left( \begin{array}{cc}
\dot{r} & E \\ E & \dot{r} \end{array}\right).
\end{equation}
We now Fourier expand the second order perturbations and the sources:
\begin{equation}
\Sigma^t(\tau,\sigma)=\sum_n\Sigma^t_n(\tau)e^{-in\sigma},\;\;\;\;\Sigma^*
(\tau,\sigma)=
\sum_n\Sigma^*_n(\tau)e^{-in\sigma},
\end{equation}
\begin{equation}
U^t(\tau,\sigma)=\sum_n U^t_n(\tau)e^{-in\sigma},\;\;\;\; U^*(\tau,\sigma)=
\sum_n U^*_n(\tau)e^{-in\sigma},
\end{equation}
and the matrix equation (3.42) reduces to:
\begin{equation}
\left( \begin{array}{c} \ddot{\Sigma}^t_n \\ \ddot{\Sigma}^*_n
\end{array}\right)+{\cal V}
\left( \begin{array}{c} \Sigma^t_n \\ \Sigma^*_n
\end{array}\right)={\cal G}^{-1}
\left( \begin{array}{c} U^t_n \\ U^*_n \end{array}\right)\equiv
\left( \begin{array}{c} \tilde{U}^t_n \\ \tilde{U}^*_n
\end{array}\right),
\end{equation}
where:
\begin{equation}
{\cal V}={\cal G}^{-1}(n^2 I+{\cal B}-{\cal A}^2-\dot{{\cal A}}){\cal G}=
\left( \begin{array}{cc} n^2+\frac{m^2}{l^2} & 0 \\ 0 & n^2
\end{array} \right),
\end{equation}
i.e. two decoupled inhomogeneous second order linear differential equations
with {\it constant} coefficients. It follows that the complete solution is
known, and given explicitly by:
\begin{eqnarray}
\Sigma^t_n(\tau)=C_n e^{-i\omega_n\tau}\hspace*{-2mm}&+&\hspace*{-2mm}
\tilde{C}_n e^{i\omega_n\tau}+\frac{e^{i\omega_n\tau}}{2i\omega_n}\int^\tau
\tilde{U}^t_n(\tau')e^{-i\omega_n\tau'}d\tau'\nonumber\\
\hspace*{-2mm}&-&\hspace*{-2mm}\frac{e^{-i\omega_n\tau}}{2i\omega_n}\int^\tau
\tilde{U}^t_n(\tau')e^{i\omega_n\tau'}d\tau'
\end{eqnarray}
and:
\begin{eqnarray}
\Sigma^*_n(\tau)=D_n e^{-in\tau}\hspace*{-2mm}&+&\hspace*{-2mm}\tilde{D}_n
e^{in\tau}+\frac{e^{in\tau}}{2in}\int^\tau\tilde{U}^*_n(\tau')e^{-in\tau'}
d\tau'\nonumber\\
\hspace*{-2mm}&-&\hspace*{-2mm}\frac{e^{-in\tau}}{2in}\int^\tau
\tilde{U}^*_n(\tau')e^{in\tau'}d\tau',
\end{eqnarray}
where:
\begin{equation}
C_n=\tilde{C}^{\dag}_
{-n},\;\;\;D_n=D^{\dag}_{-n},\;\;\;\tilde{D}_n=\tilde{D}^{\dag}_{-n}.
\end{equation}
The string perturbations $\xi^\mu$ introduced in eq. (3.4) are finally:
\begin{eqnarray}
\xi^t=\frac{1}{m(M-\frac{r^2}{l^2})}[\dot{r}\sum_n\Sigma^t_n
(\tau)e^{-in\sigma}+E\sum_n\Sigma^*_n(\tau)e^{-in\sigma}],\nonumber
\end{eqnarray}
\begin{eqnarray}
\xi^\phi=\frac{1}{r}\sum_n \Sigma^\phi_n(\tau)e^{-in\sigma},
\end{eqnarray}
\begin{eqnarray}
\xi^r=-\frac{E}{m}\sum_n\Sigma^t_n(\tau)e^{-in\sigma}-\frac{\dot{r}}{m}
\sum_n \Sigma^*_n(\tau)e^{-in\sigma}.\nonumber
\end{eqnarray}
These expressions are quite complicated because of the integrals of the sources
in eqs. (3.39) and (3.50)-(3.51). To allow comparison with the ordinary $3+1$
dimensional black hole cases (see next subsection) it is enough to consider the
region $r\rightarrow 0$ (equivalent to the region $\tau\rightarrow 0_-$).
The sources are calculated from eq. (3.16) using the first order perturbations
(3.30),(3.33). To leading order in $\tau$ for $\tau\rightarrow 0_-$ we
find:
\begin{equation}
U^\phi(\tau,\sigma)
=\frac{2E}{m(Mm^2+E^2)\;\tau^4}\sum_n\sum_p(A_{p\parallel}+
\tilde{A}_{p\parallel})(A_{n-p\perp}+\tilde{A}_{n-p\perp})e^{-in\sigma}+
{\cal O}(\frac{1}{\tau^3}),
\end{equation}
\begin{equation}
U^r(\tau,\sigma)
=\frac{M}{\sqrt{Mm^2+E^2}\;\tau^3}\sum_n\sum_p(A_{p\perp}+\tilde{A}_
{p\perp})(A_{n-p\perp}+\tilde{A}_{n-p\perp})e^{-in\sigma}+{\cal O}
(\frac{1}{\tau^2}),
\end{equation}
while $U^t$ is finite for $\tau\rightarrow 0$. It follows that:
\begin{eqnarray}
\tilde{U}^t_n(\tau)=\frac{E}{m\sqrt{Mm^2+E^2}\;\tau^3}\sum_p
(A_{p\perp}+\tilde{A}_{p\perp})(A_{n-p\perp}+\tilde{A}_{n-p\perp})+{\cal O}
(\frac{1}{\tau^2}),\nonumber
\end{eqnarray}
\begin{eqnarray}
\tilde{U}^*_n(\tau)=\frac{1}{m\;\tau^3}\sum_p(A_{p\perp}+\tilde{A}_
{p\perp})(A_{n-p\perp}+\tilde{A}_{n-p\perp})+{\cal O}(\frac{1}{\tau^2}),
\end{eqnarray}
\begin{eqnarray}
\tilde{U}^\phi_n(\tau)=\frac{-2E}{m\sqrt{Mm^2+E^2}\;\tau^3}\sum_p
(A_{p\parallel}+\tilde{A}_{p\parallel})(A_{n-p\perp}+\tilde{A}_
{n-p\perp})+{\cal O}(\frac{1}{\tau^2}).\nonumber
\end{eqnarray}
{}From eqs. (3.39) and (3.50)-(3.51) we find the asymptotic behaviour of the
$\Sigma_n$'s:
\begin{eqnarray}
\Sigma^t_n(\tau)=\frac{E}{2m\sqrt{Mm^2+E^2}\;\tau}\sum_p (A_{p\perp}+
\tilde{A}_{p\perp})(A_{n-p\perp}+\tilde{A}_{n-p\perp})+{\cal O}(1),\nonumber
\end{eqnarray}
\begin{eqnarray}
\Sigma^*_n(\tau)=\frac{1}{2m\;\tau}\sum_p(A_{p\perp}+\tilde{A}_{p\perp})
(A_{n-p\perp}+\tilde{A}_{n-p\perp})+{\cal O}(1),
\end{eqnarray}
\begin{eqnarray}
\Sigma^\phi_n(\tau)=\frac{-E}{m\sqrt{Mm^2+E^2}\;\tau}\sum_p
(A_{p\parallel}+\tilde{A}_{p\parallel})(A_{n-p\perp}+\tilde{A}_
{n-p\perp})+{\cal O}(1),\nonumber
\end{eqnarray}
and from eqs. (3.36) and (3.44)-(3.45) we have:
\begin{equation}
\xi^\phi(\tau,\sigma)=\frac{E}{mr^2}\sum_n\sum_p(A_{p\parallel}+\tilde{A}_
{p\parallel})(A_{n-p\perp}+\tilde{A}_{n-p\perp})e^{-in\sigma}+{\cal O}
(\frac{1}{r}),
\end{equation}
\begin{equation}
\xi^r(\tau,\sigma)=\frac{-M}{2r}\sum_n\sum_p(A_{p\perp}+\tilde{A}_
{p\perp})(A_{n-p\perp}+\tilde{A}_{n-p\perp})e^{-in\sigma}+{\cal O}(1),
\end{equation}
while $\xi^t$ is finite for $r\rightarrow 0\;\;(\tau\rightarrow 0_-)$. The
singularities of $\xi^\phi$ and $\xi^r$ for $r=0$ are coordinate
artifacts like the singularity of $\eta^\phi$ in eq. (3.33). Such
singularities appear even in flat Minkowski space when parametrized in terms
of polar coordinates. In the present case the coordinate singularities are
removed by introducing pseudo-Cartesian coordinates near $r=0$:
\begin{eqnarray}
T=\sqrt{M}x^t,\nonumber
\end{eqnarray}
\begin{eqnarray}
X=\frac{x^r}{\sqrt{M}}\sinh(\sqrt{M}x^\phi),
\end{eqnarray}
\begin{eqnarray}
Y=\frac{x^r}{\sqrt{M}}\cosh(\sqrt{M}x^\phi).\nonumber
\end{eqnarray}
It is easy to show that $T,\;X$ and $Y$ are finite for $r\rightarrow 0$
when $x^t,\;x^r$ and $x^\phi$ are expanded up to second order (3.4) using eqs.
(3.33) and (3.58)-(3.59).
\vskip 6pt
\hspace*{-6mm}Having
calculated the first and second order perturbations, we can now also
calculate the world-sheet energy-momentum tensor $T_{\pm\pm}$ (3.18)-(3.20).
This
calculation is simplified using the fact that $T_{\pm\pm}$ are functions of
$n(\sigma\pm\tau)$ while the first order perturbations $\eta^\mu$ are
functions of $(n\sigma\pm\omega_n\tau)$. The first order perturbations can
therefore only give constant contributions to $T_{\pm\pm}$. A straightforward
but tedious calculation gives:
\begin{eqnarray}
L_n=-2\pi i mn D_n,\;\;\;\;\tilde{L}_n=2\pi i mn
\tilde{D}_n;\;\;\;n\neq 0,\nonumber
\end{eqnarray}
\begin{eqnarray}
L_0=\pi\sum_n (\omega_n+n)^2[A_{n\parallel}\tilde{A}_{-n\parallel}+
A_{n\perp}\tilde{A}_{-n\perp}]-\frac{\pi}{2}m^2,
\end{eqnarray}
\begin{eqnarray}
\tilde{L}_0=\pi\sum_n(\omega_n-n)^2[A_{n\parallel}\tilde{A}_{-n\parallel}+
A_{n\perp}\tilde{A}_{-n\perp}]-\frac{\pi}{2}m^2,\nonumber
\end{eqnarray}
and from eq. (3.19) we get the constraints:
\begin{eqnarray}
D_n=\tilde{D}_n=0,\;\;\;\;\sum_n n\omega_n[A_{n\parallel}\tilde{A}_
{-n\parallel}+A_{n\perp}\tilde{A}_{-n\perp}]=0,\nonumber
\end{eqnarray}
as well as:
\begin{equation}
m^2=2\sum_n (2n^2+\frac{m^2}{l^2})[A_{n\parallel}
\tilde{A}_{-n\parallel}+A_{n\perp}\tilde{A}_{-n\perp}],
\end{equation}
determining the mass of the string. Notice that the mass formula of the
string is modified with respect to the usual flat space expression
$(m^2=4\sum_n n^2[A_{n\parallel}\tilde{A}_{-n\parallel}+A_{n\perp}
\tilde{A}_{-n\perp}]).$ The reason of this modification is the
asymptotic character of the spacetime, in particular the presence of a
cosmological constant. In the ordinary black hole spacetime without
$\Lambda-$term, which is asymptotically flat (and in which bounded orbits do
not appear), the mass formula is the same as in flat spacetime \cite{san2}.
In the "pure" de Sitter spacetime, however, the mass formula is modified
exactly in the same way as eq. (3.62), but for a positive $\Lambda,$ i.e.
with a term $(2n^2-m^2/l^2)$ in the sum. The quantization of the string
in the 2+1 BH-ADS background and the consequences of eq. (3.62) for
the quantum mass spectrum, are to be discussed elsewhere.

\vskip 6pt
This concludes our analysis of the first and second order perturbations
around the string center of mass, for a string embedded in the $2+1$
BH-ADS spacetime of Section 2.
\subsection{Strings in the Ordinary Black Hole Anti de Sitter Spacetime}
String perturbations around the center of mass of a string embedded in
ordinary higher dimensional black hole and de Sitter backgrounds were
already considered in Refs.[1, 10, 11]. Here we take the equatorial plane
of $3+1$ dimensional Schwarzschild anti de Sitter space, to better allow
comparison with the results of Subsection 3.2. We furthermore use the
formalism of Subsection 3.1, where only physical first order perturbations
are considered. This simplifies the analysis considerably as compared to
Refs.[1, 10, 11].

The line element is taken in the form:
\begin{equation}
ds^2=-a(r)dt^2+\frac{dr^2}{a(r)}+r^2d\theta^2+r^2\sin^2\theta d\phi^2.
\end{equation}
In the first place we take $a(r)$ to be an arbitrary function of $r$, but we
will eventually be interested in the case:
\begin{equation}
a(r)=1-\frac{2M}{r}+H^2r^2,
\end{equation}
corresponding to Schwarzschild anti de Sitter space. For a radially infalling
string in the equatorial plane $(\theta=\pi/2,\;\;\phi=\mbox{const})$
the geodesic
equations (3.3), (3.7) for the string center of mass are solved by:
\begin{equation}
\dot{t}=\frac{E}{a(r)},
\end{equation}
\begin{equation}
\dot{r}^2+m^2 a(r)=E^2,
\end{equation}
generalizing equations (3.21) and (3.22). The $\dot{r}$-equation (3.66) is
solved by:
\begin{equation}
\tau-\tau_o=\int_{r_o}^r\frac{dx}{\sqrt{E^2-m^2 a(x)}},
\end{equation}
and after inversion, giving $r$ as an explicit function of $\tau$, the
coordinate time $t$ is obtained by integration of eq. (3.65). In the case of a
Schwarzschild black hole this gives the well-known results in terms of
elementary functions (see for instance Ref.\cite{mis}, \S 25.5), while in the
case under consideration here, where the function $a(r)$ is given by eq.
(3.64), $r(\tau)$ and $t(\tau)$ will be expressed in terms of elliptic
functions.

There are now three covariantly constant normal vectors fulfilling eqs. (3.8)
and (3.10):
\begin{eqnarray}
n^\mu_{1\perp}=(0,\;0\;\frac{1}{r},\;0),\nonumber
\end{eqnarray}
\begin{eqnarray}
n^\mu_{2\perp}=(0,\;0,\;0,\;\frac{1}{r}),
\end{eqnarray}
\begin{eqnarray}
n^\mu_\parallel=(\frac{\dot{r}}{m a(r)},\;\frac{E}{m},\;0,\;0).\nonumber
\end{eqnarray}
The non-vanishing components of the Riemann tensor, corresponding to the
line element (3.63), are:
\begin{eqnarray}
&R_{rtrt}=\frac{1}{2}a_{,rr}\;\;R_{r\theta r\theta}=\frac{-r}{2a}a_{,r},\;\;
R_{r\phi r\phi}=\frac{-r}{2a}a_{,r}\sin^2\theta,&\nonumber\\
&R_{t\theta t\theta}=\frac{r}{2}aa_{,r},\;\;R_{t\phi t\phi}=\frac{r}{2}
aa_{,r}\sin^2\theta,\;\;R_{\theta\phi\theta\phi}=r^2(1-a)\sin^2\theta.&
\end{eqnarray}
Equations (3.14), determining the Fourier components of the comoving first
order perturbations, separate and reduce to:
\begin{equation}
\ddot{C}_{nS\perp}+(n^2+\frac{m^2 a_{,r}}{2r})
C_{nS\perp}=0;\;\;\;S=1,2
\end{equation}
\begin{equation}
\ddot{C}_{n\parallel}+(n^2+\frac{m^2 a_{,rr}}{2})C_{n\parallel}=0.
\end{equation}
For the two transverse polarizations we find from eq. (3.64):
\begin{equation}
\ddot{C}_{nS\perp}+(n^2+m^2 H^2+\frac{Mm^2}{r^3})
C_{nS\perp}=0,\;\;\;S=1,2
\end{equation}
while for the longitudinal polarization:
\begin{equation}
\ddot{C}_{n\parallel}+(n^2+m^2 H^2-\frac{2Mm^2}{r^3})
C_{n\parallel}=0.
\end{equation}
{}From these equations it is obvious that we need only look for singularities
in the region $r\rightarrow 0$. However, for the transverse perturbations
(3.72) the bracket is always positive. We therefore expect that the solution
is oscillating and bounded even for $r\rightarrow 0$. For the longitudinal
perturbations (3.73), on the other hand, the bracket can be negative. In that
case imaginary frequencies arise and instabilities develop.
The $(\mid n\mid=1)$-instability sets in at:
\begin{equation}
r_{\mbox{inst.}}=(\frac{2Mm^2}{1+m^2 H^2})^{\frac{1}{3}},
\end{equation}
which may be inside or outside the horizon, depending on the relation between
the various parameters $(M,H,m)$ involved. The
higher modes develop instabilities for
smaller $r$.

These results are easily confirmed by the exact time evolution of the
perturbations. For $r\rightarrow 0$ we find from eq.
(3.66):
\begin{equation}
r(\tau)\approx(3m\sqrt{M/2})^{\frac{2}{3}}(\tau_0-\tau)^
{\frac{2}{3}},
\end{equation}
where the integration constant is chosen such that $r\rightarrow 0$
corresponds to $\tau\rightarrow\tau_0.$
Then for $\tau\rightarrow\tau_0$ eqs. (3.72) and (3.73) are:
\begin{equation}
\ddot{C}_{nS\perp}+\frac{2}{9(\tau_0-\tau)^2}C_{nS\perp}=0;\;\;\;S=1,2
\end{equation}
\begin{equation}
\ddot{C}_{n\parallel}-\frac{4}{9(\tau_0-\tau)^2}C_{n\parallel}=0,
\end{equation}
with complete solutions:
\begin{equation}
C_{nS\perp}(\tau)=\alpha_{nS\perp}(\tau_0-\tau)^{\frac{1}{3}}+\beta_{nS\perp}
(\tau_0-\tau)^{\frac{2}{3}};\;\;\;S=1,2
\end{equation}
\begin{equation}
C_{n\parallel}(\tau)=\gamma_{n\parallel}(\tau_0-\tau)^{\frac{4}{3}}
+\delta_{n\parallel}
(\tau_0-\tau)^{-\frac{1}{3}},
\end{equation}
where $(\alpha_{nS\perp},\;\beta_{nS\perp},\;\gamma_{n\parallel},\;
\delta_{n\parallel})$ are constants. It follows that $C_{nS\perp}(\tau)$
is finite for $r\rightarrow 0\;\;(\tau\rightarrow \tau_0)$ while
$C_{n\parallel}(\tau)$ blows up because of the $\delta_{n\parallel}$-term.
This result demonstrates the important difference between the $2+1$ BH-ADS
solution of Section 2 and the equatorial plane of an ordinary higher
dimensional black hole. In the first case we found finite bounded
perturbations, while in the latter case instabilities develop already in the
first order comoving perturbations near $r=0$. For the ordinary black hole
anti de Sitter spacetime it is then meaningless to analyze higher order
perturbations for $r\rightarrow 0,$ in contrast, as we have seen, to
the $2+1$ BH-ADS background. In the ordinary black hole anti de Sitter
spacetime the string falls to the center $(r=0)$ and is trapped by the
singularity. For $\tau\rightarrow \tau_0$ the potential for the radial
perturbations,
eq. (3.77), is of the type $-\gamma/(\tau-\tau_0)^2$ with $\gamma=4/9,$ which
is a singular attractive potential.
\subsection{Strings in the Black String Background}
We close this section with the analysis of the string propagation
in the black string background. It is convenient
to take the metric in the form (2.14) and for simplicity we consider the
uncharged $({\cal Q}=0)$ black string:
\begin{equation}
ds^2=-(1-\frac{{\cal M}}{r})dt^2+(1-\frac{{\cal M}}{r})^{-1}\frac{l^2
dr^2}{4r^2}+dx^2,
\end{equation}
where the tildes have been deleted (c.f. eq. (2.14)). This spacetime is
just the direct product of Witten's $2$ dimensional black hole \cite{wit} and
the real line space. It has a horizon at $r={\cal M}=Ml$ and, contrary to its
dual, the $2+1$ BH-ADS solution, it has a strong
curvature singularity at $r=0.$ To
compare with the results of Subsections 3.2 and 3.3 we consider a
radially infalling string, corresponding to $x=\mbox{const.}$
Equations (3.3) and
(3.7) for the string center of mass become:
\begin{equation}
\dot{t}=\frac{E}{1-{\cal M}/r},
\end{equation}
\begin{equation}
\dot{r}^2=\frac{4r^2}{l^2}[E^2-m^2+\frac{m^2{\cal M}}{r}],
\end{equation}
which can be solved in terms of elementary functions (taking for simplicity
$m^2>E^2$):
\begin{equation}
r(\tau)=\frac{m^2{\cal M}}{2(m^2-E^2)}[1-\sin\frac{2\sqrt{m^2-E^2}}{l}\tau],
\end{equation}
and:
\begin{equation}
t(\tau)=E\tau-\frac{l}{2}\log\mid\frac{1+\frac{m^2-2E^2}{m^2-2E\sqrt{m^2-E^2}}
\tan\frac{\sqrt{m^2-E^2}}{l}\tau}{1+\frac{m^2-2E^2}{m^2+2E\sqrt{m^2-E^2}}
\tan\frac
{\sqrt{m^2-E^2}}{l}\tau}\mid,
\end{equation}
where the integration constants were chosen such that $t(0)=0$ and
$r(\tau_0)=0$ for:
\begin{equation}
\tau_0=\frac{\pi l}{4\sqrt{m^2-E^2}}.
\end{equation}
Notice that the horizon is passed for $\tau=\tau_{\mbox{hor}}$:
\begin{equation}
\tau_{\mbox{hor}}=\frac{l}{2\sqrt{m^2-E^2}}\arcsin(\frac{2E^2}{m^2}-1),
\end{equation}
and that $t(\tau_{\mbox{hor}})=\infty.$

The two covariantly constant normal vectors, fulfilling eqs. (3.8) and
(3.10), are given by:
\begin{eqnarray}
n^\mu_\perp=(0,\;0,\;1),\nonumber
\end{eqnarray}
\begin{eqnarray}
n^\mu_\parallel=(\frac{l\dot{r}}{2m(r-{\cal M})},\;\frac{2Er}{ml},\;0).
\end{eqnarray}
The only non-vanishing component of the Riemann tensor, corresponding to the
line element (3.80), is:
\begin{equation}
R_{trtr}=\frac{-{\cal M}}{2r^3},
\end{equation}
and then equations (3.14), determining the string perturbations, take the form:
\begin{equation}
\ddot{C}_{n\perp}+n^2 C_{n\perp}=0,
\end{equation}
\begin{equation}
\ddot{C}_{n\parallel}+(n^2-\frac{2m^2{\cal M}}{l^2 r})C_{n\parallel}=0.
\end{equation}
Not surprisingly, the perturbations in the $x$-direction are completely finite
and regular. For the longitudinal perturbations we see that the bracket in
eq. (3.90) becomes negative and approaches $-\infty$ for $r\rightarrow 0,$
suggesting an instability. This is confirmed by the exact time evolution near
$r=0.$ Using eq. (3.82) we find for $r\rightarrow 0$:
\begin{equation}
r(\tau)\approx\frac{m^2{\cal M}}{l^2}(\tau_0-\tau)^2,
\end{equation}
where the integration constant is chosen such that $r\rightarrow 0$
corresponds to $\tau\rightarrow\tau_0.$ Equation (3.90) is now for
$\tau\rightarrow\tau_0:$
\begin{equation}
\ddot{C}_{n\parallel}-\frac{2}{(\tau_0-\tau)^2}C_{n\parallel}=0,
\end{equation} with solution:
\begin{equation}
C_{n\parallel}(\tau)=\alpha_{n\parallel}(\tau_0-\tau)^2+\frac
{\beta_{n\parallel}}{\tau_0-\tau}.
\end{equation}
The solution indeed blows up for $r\rightarrow 0\;$ ($\tau\rightarrow
\tau_0$) with conclusions similar to the ordinary black hole case, Subsection
3.3.
\section{Circular Strings in Stationary Axially Symmetric Backgrounds}
\setcounter{equation}{0}
In this section we consider circular strings embedded in stationary axially
symmetric backgrounds. Circular strings in curved spacetimes have attracted
important interest recently [6, 8, 32-38]. The
analysis is carried out in $2+1$ dimensions, but
the results will hold for the equatorial plane of higher dimensional
backgrounds as well. To be more specific we consider the following line
element:
\begin{equation}
ds^2=g_{tt}(r)dt^2+g_{rr}(r)dr^2+2g_{t\phi}(r)dtd\phi+g_{\phi\phi}(r)d\phi^2,
\end{equation}
that will be general enough for our purposes here. It obviously includes
as special cases the BH-ADS solution of Section 2 (as well as the black
string) and the equatorial plane of the black hole solutions of Einstein theory
in $3+1$ dimensions.

The circular string ansatz, consistent with the symmetries of the background,
is taken to be:
\begin{equation}
t=t(\tau),\;\;\;r=r(\tau),\;\;\;\phi=\sigma+f(\tau),
\end{equation}
where the three functions $t(\tau),\;r(\tau)$ and
$f(\tau)$ are to be determined
by the equations of motion and constraints (3.1)-(3.2). The equations of motion
(3.1) for the ansatz (4.2) and the background (4.1) lead to:
\begin{eqnarray}
\ddot{t}+2\Gamma^t_{tr}\dot{t}\dot{r}+2\Gamma^t_{\phi r}\dot{r}
\dot{f}=0,\nonumber
\end{eqnarray}
\begin{eqnarray}
\ddot{r}+\Gamma^r_{rr}\dot{r}^2+\Gamma^r_{tt}\dot{t}^2+\Gamma^r_{\phi\phi}
(\dot{f}^2-1)+2\Gamma^r_{t\phi}\dot{t}\dot{f}=0,
\end{eqnarray}
\begin{eqnarray}
\ddot{f}+2\Gamma^\phi_{tr}\dot{t}\dot{r}+2\Gamma^\phi_{\phi r}\dot{f}
\dot{r}=0,\nonumber
\end{eqnarray}
while the constraints become:
\begin{eqnarray}
g_{tt}\dot{t}^2+g_{rr}\dot{r}^2+g_{\phi\phi}(\dot{f}^2+1)+2g_{t\phi}
\dot{t}\dot{f}=0,\nonumber
\end{eqnarray}
\begin{eqnarray}
g_{t\phi}\dot{t}+g_{\phi\phi}\dot{f}=0.
\end{eqnarray}
This system of second order ordinary differential equations and constraints is
most easily described as a Hamiltonian system:
\begin{equation}
{\cal H}=\frac{1}{2}g^{tt}P^2_t+\frac{1}{2}g^{rr}P^2_r+\frac{1}{2}
g^{\phi\phi}P^2_\phi+g^{t\phi}P_tP_\phi+\frac{1}{2}g_{\phi\phi},
\end{equation}
supplemented by the constraints:
\begin{equation}
{\cal H}=0,\;\;\;\;P_\phi=0.
\end{equation}
The function $f(\tau)$ introduced in eq. (4.2) does not represent any physical
degrees of freedom. It describes the "longitudinal" rotation of the circular
string and is therefore a pure gauge artifact. This interpretation is
consistent with eq. (4.6) saying that there is no angular momemtum $P_\phi$.

The Hamilton equations of the two cyclic coordinates $t$ and $f$ are:
\begin{equation}
\dot{f}=g^{\phi\phi}P_\phi+g^{t\phi}P_t,\;\;\;\;
\dot{t}=g^{tt}P_t+g^{t\phi}P_\phi,
\end{equation}
as well as:
\begin{equation}
P_\phi=\mbox{const.}=0,\;\;\;\;P_t=\mbox{const.}\equiv -E,
\end{equation}
where $E$ is an integration constant and we used eq. (4.6). The two functions
$t(\tau)$ and $f(\tau)$ are then determined by:
\begin{equation}
\dot{f}=-Eg^{t\phi},
\end{equation}
\begin{equation}
\dot{t}=-Eg^{tt},
\end{equation}
that can be integrated provided $r(\tau)$ is known. Using eqs. (4.6) and
(4.9)-(4.10) the Hamilton equation of $r$ becomes after one integration:
\begin{equation}
\dot{r}^2=-g^{rr}(E^2g^{tt}+g_{\phi\phi}),
\end{equation}
so that $r(\tau)$ can be obtained by inversion of:
\begin{equation}
\tau-\tau_o=\int_{r_o}^r\frac{dx}{\sqrt{-g^{rr}(x)[E^2g^{tt}(x)+
g_{\phi\phi}(x)]}}.
\end{equation}
For the cases that we will consider in the following, eq. (4.12) will be solved
in terms of either elementary or elliptic functions.
\vskip 6pt
\hspace*{-8mm}
We close this subsection by the following interesting observation: Insertion
of the ansatz (4.2), using the results (4.9)-(4.11), in the line element (4.1)
leads to:
\begin{equation}
ds^2=g_{\phi\phi}(d\sigma^2-d\tau^2).
\end{equation}
We can then identify the invariant string size as:
\begin{equation}
S(\tau)=\sqrt{g_{\phi\phi}(r(\tau))}
\end{equation}
For the $2+1$ dimensional BH-ADS spacetime
the invariant string size is then simply $r,$ as well as for the
ordinary $3+1$ dimensional Schwarzschild and
Reissner-Nordstr\"{o}m black hole backgrounds.
For the black string background (2.13)
the invariant string size is actually $r^{-1}.$
\subsection{Circular Strings in the 2+1 BH-ADS Spacetime}
In the $2+1$ dimensional BH-ADS spacetime (2.6), equation (4.11) determining
the invariant string size $r(\tau)$, takes the explicit form:
\begin{equation}
\dot{r}^2+V(r)=0;\;\;\;\;V(r)=r^2(\frac{r^2}{l^2}-M)+\frac{J^2}{4}-E^2.
\end{equation}
Here we have defined the potential $V(r)$ such that the dynamics takes place at
the $r$-axis in a $(r,V(r))$ diagram, see Fig.1. The potential (4.15) has a
global minimum between the two horizons:
\begin{equation}
V_{\mbox{min}}=V(\sqrt{\frac{Ml^2}{2}}\;)=-\frac{1}{4}(M^2l^2-J^2+4E^2)<0,
\end{equation}
which is always negative, since we only consider the case when $Ml^2\geq J^2$
(otherwise there are no horizons, see eq. (2.7)). For large values of $r$ the
potential goes as $r^4$ and at $r=0$ we have:
\begin{equation}
V(0)=\frac{J^2}{4}-E^2,
\end{equation}
that can be either positive, negative or zero. Notice also that the potential
vanishes provided:
\begin{equation}
V(r_0)=0\;\;\Leftrightarrow\;\;r_{01,2}=\sqrt{\frac{Ml^2}{2}\pm
\frac{l}{2}\sqrt{M^2l^2-J^2+4E^2}}.
\end{equation}
Equation (4.18) leads to three fundamentally different types of solutions.
\vskip 6pt
\hspace*{-6mm}{\bf (i)}: For $J^2>4E^2$ there are two positive-$r$ zeroes
of the
potential (Fig.1a). The smallest zero is located between the inner horizon
and $r=0$, while the other zero is between the outer horizon and the static
limit. Therefore, this string solution never comes outside the static limit. On
the other hand it never falls into $r=0$. The mathematical solution
oscillating between these two positive zeroes of the potential may be
interpreted as a string travelling between the different universes described by
the maximal analytic extension of the spacetime (2.6) (the Penrose diagram of
the $2+1$ dimensional BH-ADS spacetime is discussed in Refs.[23, 25]. Such
type of
circular string solutions also exist in other stringy
black hole backgrounds \cite{veg3}.
\vskip 6pt
\hspace*{-6mm}{\bf (ii)}: For $J^2<4E^2$ there is only one positive-$r$
zero of the potential, which
is always located outside the static limit (Fig.1b). The
potential is negative for $r=0$, so there is no barrier preventing the string
from collapsing into $r=0$. By suitably fixing the initial conditions the
string starts with its maximal size outside the static limit at $\tau=0$. It
then contracts through the ergosphere and the two horizons and eventually
falls into $r=0$. If $J\neq 0$ it may however still be possible to continue
this solution into another universe as in the case {\bf (i)}.
\vskip 6pt
\hspace*{-6mm}{\bf (iii)}: $J^2=4E^2$ is the limiting case where the maximal
string radius equals the static limit. The potential is exactly zero for
$r=0$ so also in this case the string contracts through the two horizons and
eventually falls into $r=0$.
\vskip 6pt
\hspace*{-6mm}
Let us now look at the exact mathematical solution of eq. (4.15). In the
general case (arbitrary $J$) the non-negative solution can be represented as:
\begin{equation}
r(\tau)=\mid r_m-\frac{1}{c_1\wp(\tau-\tau_o)+c_2}\mid,
\end{equation}
where $r_m$ is the maximal string radius:
\begin{equation}
r_m=\sqrt{\frac{Ml^2}{2}+\frac{l}{2}\sqrt{M^2l^2-J^2+4E^2}}\;,
\end{equation}
$c_1$ and $c_2$ are two constants given by:
\begin{equation}
c_1=(\frac{r^3_m}{l^2}-\frac{Mr_m}{2})^{-1},\;\;\;\;c_2=\frac{1}{12}
(\frac{6r^2_m}{l^2}-M)(\frac{r^3_m}{l^2}-\frac{Mr_m}{2})^{-1},
\end{equation}
and $\wp$ is the Weierstrass elliptic $\wp$-function \cite{abr}:
\begin{equation}
\dot{\wp}^2=4\wp^3-g_2\wp-g_3,
\end{equation}
with invariants:
\begin{equation}
g_2=\frac{M^2}{12}+\frac{Mr^2_m}{l^2}-\frac{r^4_m}{l^4},\;\;\;\;g_3=
-\frac{M^3}{216}+\frac{M^2r^2_m}{6l^2}-\frac{Mr^4_m}{6l^4},
\end{equation}
discriminant:
\begin{equation}
\Delta\equiv g^3_2-27g^2_3=\frac{r^2_m}{l^2}(M-\frac{r^2_m}{l^2})
(\frac{M}{2}-\frac{r^2_m}{l^2})^4,
\end{equation}
and roots:
\begin{equation}
\left\{ -\frac{M}{6},\;\;\;\frac{1}{2}(\frac{M}{6}\pm\frac{r_m}{l^2}\sqrt
{Ml^2-r^2_m}\;)\right\}.
\end{equation}
The qualitatively different solutions {\bf (i)},{\bf (ii)} and {\bf (iii)}
discussed before, are now distinguished by the sign of the discriminant
$\Delta$:
\begin{equation}
J^2-4E^2>(<)\;0\;\;\Leftrightarrow\;\;r_m^2<(>)\;Ml^2\;\;\Leftrightarrow\;\;
\Delta>(<)\;0,
\end{equation}
where we used that $Ml^2<2r^2_m$ (otherwise there are no horizons). We then
analyze the three types of solutions in terms of the mathematical formalism
above.
\vskip 6pt
\hspace*{-6mm}{\bf (i)}, $\Delta>0$: In this case the roots (4.25) are given
by:
\begin{equation}
e_1\equiv\frac{M}{12}+\frac{r_m}{2l^2}\sqrt{Ml^2-r^2_m}\;>\;0\;\geq e_2
\equiv\frac{M}{12}-\frac{r_m}{2l^2}\sqrt{Ml^2-r^2_m}\;>\;e_3\equiv-\frac{M}{6},
\end{equation}
and the solution (4.19) can be written in terms of Jacobian elliptic functions:
\begin{equation}
r(\tau)=r_m\frac{\delta-\mbox{sn}^2[\tau^*,k]}{\delta+\mbox{sn}^2[\tau^*,k]},
\end{equation}
where:
\begin{equation}
\tau^*=\sqrt{e_1-e_3}\;\tau,\;\;\;\;\delta=\frac{4l^2(e_1-e_3)}
{2r^2_m-Ml^2},\;\;\;\;
k=\sqrt{\frac{e_2-e_3}{e_1-e_3}}.
\end{equation}
It follows that:
\begin{equation}
r(0)=r_m,\;\;\;\;r(\omega)=\sqrt{Ml^2-r^2_m},\;\;\;\;r(2\omega)=r_m,\;....
\end{equation}
where $\omega$ is the real semi-period of the Weierstrass function:
\begin{equation}
\omega=\frac{K(k)}{\sqrt{e_1-e_3}},
\end{equation}
and $K(k)$ is the complete elliptic integral of first kind. From eqs. (4.28),
(4.30)
is seen that the solution oscillates between the two positive zeroes (4.18) of
the potential, with the period $2\omega$.
\vskip 6pt
\hspace*{-6mm}{\bf (ii)}, $\Delta<0$: Now two of the roots (4.25) become
non-real:
\begin{equation}
e_1\equiv\frac{M}{12}+i\frac{r_m}{2l^2}\sqrt{r_m^2-Ml^2},\;\;\;e_3\equiv
\frac{M}{12}-i\frac{r_m}{2l^2}\sqrt{r^2_m-Ml^2}\;\;\;e_2\equiv-\frac{M}
{6},
\end{equation}
and eq. (4.19) leads to:
\begin{equation}
r(\tau)=r_m\mid \mbox{cn}[2\sqrt{H_2}\;\tau,k]\mid,
\end{equation}
where:
\begin{equation}
H_2=\sqrt{e^2_2-e_1 e_3},\;\;\;k=\sqrt{\frac{1}{2}-\frac{3e_2}{4H_2}}.
\end{equation}
It follows that:
\begin{equation}
r(0)=r_m,\;\;\;\;r(\frac{\omega_2}{2})=0,\;\;\;\;r(\omega_2)=r_m,\;....
\end{equation}
where the real semi-period of the Weierstrass function is now:
\begin{equation}
\omega_2=\frac{K(k)}{\sqrt{H_2}}.
\end{equation}
\vskip 6pt
\hspace*{-6mm}{\bf (iii)}, $\Delta=0$: In this limiting case the elliptic
functions reduce to hyperbolic functions. Explicitly we find:
\begin{equation}
r(\tau)=\frac{\sqrt{M}l}{\cosh(\sqrt{M}\tau)},
\end{equation}
so that:
\begin{equation}
r(-\infty)=0,\;\;\;\;r(0)=r_m=\sqrt{M}l,\;\;\;\;r(\infty)=0.
\end{equation}
In this limiting case $\Delta=0$ the mathematical solution only makes one
oscillation between $r=0$ and the maximal radius.
\vskip 6pt
\hspace*{-6mm}Let us close this section with a few more words on the
non-rotating black hole case. For $J=0$ we are always in case {\bf (ii)}, i.e.
if the black hole has no angular momentum the circular string has its maximal
invariant size larger than the horizon (there is no ergosphere and only one
horizon in this case) and it always falls into $r=0$. The Penrose diagram
for $J=0$ \cite{ban2} is similar to the Penrose diagram of ordinary
Schwarzschild spacetime, so the
string motion outwards from $r=0$ is unphysical because of the causal
structure. The string motion stops when the string falls into
$r=0$.

The physical string size is given by eq. (4.33) for $J=0$, and the coordinate
time $t(\tau)$ is then obtained from eq. (4.10):
\begin{equation}
t(\tau)=-El^2\int^\tau\frac{dx}{Ml^2-r^2_m \mbox{cn}^2[2\sqrt{H_2}\;x,k]},
\end{equation}
that leads to:
\begin{equation}
t(\tau)=\frac{El^2}{2(r^2_m-Ml^2)\sqrt{H_2}}\Pi\left(
\frac{r^2_m}{r^2_m-Ml^2},\;\sqrt{H_2}\tau,\;k\right).
\end{equation}
Here $\Pi$ is the incomplete elliptic integral of third kind
and $r_m$, $k,$ $H_2$
and $\omega_2$ are given by eqs. (4.20), (4.34) and (4.36), respectively, with
$J=0$. The string has its maximal size for $\tau=0$, passes the horizon
$r_{\mbox{hor}}=\sqrt{M}l$ at:
\begin{equation}
\tau_{\mbox{hor}}=\frac{1}{2\sqrt{H_2}}F\left( \arcsin\sqrt{1-\frac{Ml^2}
{r^2_m}},\;k\right),
\end{equation}
where $F$ is the incomplete elliptic integral of first kind, and falls into
$r=0$ for $\tau=\omega_2/2$, (eq. (4.35)). The corresponding values of the
coordinate time are given by:
\begin{equation}
t(0)=0,\;\;t(\tau_{\mbox{hor}})=\infty,\;\;t(\frac{\omega_2}{2})=
\frac{Elr_m}{2\sqrt{MH_2}\sqrt{r^2_m-Ml^2}}\frac{Z[\epsilon,k]}
{\sqrt{r^2_m(1-k^2)+k^2Ml^2}},
\end{equation}
where $Z[\epsilon,k]$ is the Jacobian zeta function \cite{abr}
and $\mbox{sn}[\epsilon,k]=(r^2_m-Ml^2)/r^2_m.$
\subsection{Circular Strings in Ordinary Spacetimes}
We will now compare the circular strings in the
$2+1$ dimensional BH-ADS spacetime and in the equatorial plane
of ordinary $3+1$
dimensional black holes. In the most general case it is natural
to compare the spacetime
metric (2.6) to the ordinary $3+1$ dimensional Kerr anti de
Sitter spacetime with metric components:
\begin{eqnarray}
g_{tt}=\frac{a^2\Delta_\theta\sin^2\theta-\Delta_r}{\rho^2},\;\;\;\;g_{rr}=
\frac{\rho^2}{\Delta_r},\;\;\;\;g_{t\phi}=(\Delta_r-(r^2+a^2)\Delta_\theta)
\frac{a\sin^2\theta}{\Delta_o\rho^2},\nonumber
\end{eqnarray}
\begin{eqnarray}
g_{\phi\phi}=\left( \Delta_\theta(r^2+a^2)^2-a^2\Delta_r\sin^2\theta\right)
\frac{\sin^2\theta}{\Delta^2_o\rho^2},\;\;\;\;g_{\theta\theta}=\frac{\rho^2}
{\Delta_\theta},
\end{eqnarray}
where we have introduced the notation:
\begin{eqnarray}
&\Delta_r=(1-\frac{1}{3}\Lambda r^2)(r^2+a^2)-2Mr,\;\;\;\;\Delta_\theta=
1+\frac{1}{3}\Lambda a^2\cos^2\theta,&\nonumber\\
&\Delta_o=1+\frac{1}{3}\Lambda a^2,\;\;\;\;\rho^2=r^2+a^2\cos^2\theta.&
\end{eqnarray}
Here the mass is represented by $M$ while $a$ is the specific angular momentum,
and a positive $\Lambda$ corresponds to de Sitter while a negative $\Lambda$
corresponds to anti de Sitter spacetime. In
the equatorial plane $(\theta=\pi/2)$ the
metric (4.43) is in the general form (4.1) and it is easy to see that the
analysis of Section 4 goes through, so that we can take over the general
results of eqs. (4.9),(4.10) and (4.11) for the circular strings. In the
most general case the $t$-equation (4.10) and the
$r$-equation (4.11)
take the form:
\begin{equation}
\dot{t}=\frac{E}{\Delta_r r^2}[(r^2+a^2)^2-a^2\Delta_r],\;\;\;
\;\;\dot{r}^2+V(r)=0,
\end{equation}
where the potential is given explicitly by:
\begin{eqnarray}
V(r)\hspace*{-2mm}&=&\hspace*{-2mm}-\frac
{\Lambda}{3\Delta_o}r^4+\frac{1-2\Lambda a^2/3}{\Delta_o}r^2-
\frac{2M(1+2\Lambda a^2/3)}{\Delta^2_o}r\nonumber\\
\hspace*{-2mm}&+&\hspace*{-2mm}\frac{2a^2-\Lambda a^2/3-E^2
\Delta^2_o}{\Delta_o}
-\frac{4M\Lambda a^4}{3\Delta^2_o}\frac{1}{r}\nonumber\\
\hspace*{-2mm}&+&\hspace*{-2mm}\frac{a^2(\Delta_o(a^2-E^2\Delta^2_o)
-4M^2)}{\Delta^2_o}\frac{1}{r^2}+
\frac{2Ma^2(a^2-E^2\Delta^2_o)}{\Delta^2_o}\frac{1}{r^3}
\end{eqnarray}
i.e. the potential covers seven powers in $r$. The general
solution will therefore
involve higher genus elliptic functions and we shall not go into any details
here. It is furthermore very complicated to deduce the physical properties of
the circular strings from the shape of the potential (the zeroes etc.)
since the invariant string size (4.14) is non-trivially connected to $r:$
\begin{equation}
S(\tau)=\sqrt{\frac{r^2+a^2}{1+\Lambda a^2/3}+\frac{2M}{r}
\frac{a^2}{(1+\Lambda a^2/3)^2}}.
\end{equation}

We will leave the general Kerr anti de Sitter case for analysis elsewhere, and
concentrate here
on the non-rotating case $a=0$, that should be compared with the
$J=0$ case of the $2+1$ BH-ADS solution discussed at the end of Subsection 4.1.
Our strategy will be to start with the simplest case of a circular string in
flat Minkowski spacetime
and then introduce a mass (Schwarzschild) and
a negative cosmological constant (anti de Sitter).
\vskip 6pt
\hspace*{-6mm}Minkowski space:\\
This case was originally discussed by
Vilenkin \cite{vil4}, and is obtained
from eqs. (4.45)-(4.46) taking $\Lambda=0=a=M$:
\begin{equation}
\dot{t}=E\;\;\;\;\;V(r)=r^2-E^2,
\end{equation}
i.e. (see Fig.2a):
\begin{equation}
V(0)=-E^2,\;\;\;\;r_m=E,\;\;\;\;V(r)\propto r^2\;\;\;\mbox{for}\;\;\;r>>E.
\end{equation}
The string oscillates between its maximal size
$r=E$ and $r=0$, with the solution of eqs. (4.45)
given explicitly by:
\begin{equation}
r(\tau)=r_m\mid\cos\tau\mid,\;\;\;\;\;t=E\tau.
\end{equation}
\vskip 6pt
\hspace*{-6mm}Schwarzschild black hole:\\
This case was already considered in Ref.\cite{veg3} for $E=0$ and in
Ref.\cite{all2} for arbitrary $E$. The potential and coordinate time are
obtained from eqs. (4.45)-(4.46)
with $a=0=\Lambda$:
\begin{equation}
\dot{t}=\frac{E}{1-2M/r},\;\;\;\;\;V(r)=r^2-2Mr-E^2,
\end{equation}
i.e. (see Fig.2b):
\begin{equation}
V(0)=-E^2,\;\;\;\;r_m=M+\sqrt{M^2+E^2}>E,\;\;\;\;V(r)
\propto r^2\;\;\mbox{for}\;\;r>>(M,E).
\end{equation}
The mathematical solution oscillates between $r=M+\sqrt{M^2+E^2}$ and $r=M-
\sqrt{M^2+E^2}<0$,
but because of the causal structure and the curvature singularity the motion
stops at $r=0.$ The solution of eqs. (4.45) is remarkably
simple (compare with the point particle case, see for instance Ref.\cite{mis}):
\begin{eqnarray}
r(\tau)=M+\sqrt{M^2+E^2}\;\cos\tau,\nonumber
\end{eqnarray}
\begin{eqnarray}
t(\tau)=E\tau+2M\log\mid\frac{\tan\frac{\tau}{2}+(\sqrt{M^2+E^2}-M)/E}
{\tan\frac{\tau}{2}-(\sqrt{M^2+E^2}-M)/E}\mid.
\end{eqnarray}
\vskip 6pt
\hspace*{-6mm}Anti de Sitter space:\\
Here we take $a=0=M$ and $\Lambda\equiv-3H^2<0$ in eqs. (4.45)-(4.46) and find:
\begin{equation}
\dot{t}=\frac{E}{1+H^2r^2},\;\;\;\;\;V(r)=r^2(1+H^2r^2)-E^2,
\end{equation}
i.e. (see Fig.2c):
\begin{eqnarray}
&V(0)=-E^2,\;\;\;\;r_m=\frac{1}{\sqrt{2}H}\sqrt{-1+\sqrt
{1+4H^2E^2}}<E,&\nonumber\\
&V(r)\propto r^4\;\;\mbox{for}\;\;r>>(1/H,E).&
\end{eqnarray}
The string is oscillating between $r=\frac{1}{\sqrt{2}H}
\sqrt{-1+\sqrt{1+4H^2E^2}}$ and $r=0$. The solution of eqs. (4.45)
for $r(\tau)$ is:
\begin{equation}
r(\tau)=r_m\mid \mbox{cn}[(1+4H^2E^2)^{\frac{1}{4}}\tau,k]\mid,
\end{equation}
which is periodic with period $2\omega$:
\begin{equation}
\omega=\frac{K(k)}{(1+4H^2E^2)^{\frac{1}{4}}};\;\;\;k=\sqrt{\frac{\sqrt{1+
4H^2E^2}-1}{2\sqrt{1+4H^2E^2}}}.
\end{equation}
Equation (4.54) can then be integrated, and we obtain
for $t(\tau)$:
\begin{equation}
t(\tau)=\frac{E}{(1+4H^2E^2)^{\frac{1}{4}}
(1+H^2r_m^2)}\Pi\left( \frac{H^2r^2_m}
{1+H^2r_m^2},\;(1+4H^2E^2)^{\frac{1}{4}}\tau,\;k\right),
\end{equation}
where $\Pi$ is the incomplete elliptic integral of the third kind.
\vskip 6pt
\hspace*{-6mm}Schwarzschild anti de Sitter space:\\
By taking $a=0$ and $\Lambda=-3H^2<0$ in eqs. (4.45)-(4.46)
we are finally in the case of
Schwarzschild anti de Sitter spacetime:
\begin{equation}
\dot{t}=\frac{E}{1+H^2r^2-2M/r},\;\;\;\;\;V(r)=H^2r^4+r^2-2Mr-E^2,
\end{equation}
i.e. (see Fig.2d):
\begin{eqnarray}
&V(0)=-E^2,\;\;\;\;H^2r^4_m+r^2_m-2Mr_m-E^2=0,&\nonumber\\
&V(r)
\propto r^4\;\;\;\mbox{for}\;\;\;r>>
(M,H^{-1},E).&
\end{eqnarray}
Notice that $V(r)\leq -E^2$ inside the horizon, and that $dV/dr$ has only one
real (positive) zero. It follows that the $r_m$-equation (4.60) has exactly
one positive solution which is then by definition $r_m$. The explicit
(but not very enlightening)
expression for $r_m$ as a function of $M$, $H$ and $E$ can of course be written
down by solving the quartic equation, but we shall not give the result here.
The solution of
eqs. (4.45) for $r(\tau)$
can be written in terms of the Weierstrass elliptic $\wp$-function:
\begin{equation}
r(\tau)=r_m-\frac{1}{d_1\wp(\tau-\tau_o)+d_2},
\end{equation}
where the two constants $d_1$ and $d_2$ are given by:
\begin{equation}
d_1=2(r_m-M+2H^2r^3_m)^{-1},\;\;\;\;d_2=\frac{1}{6}(1+6H^2r^2_m)(r_m-M+2H^2
r^3_m)^{-1}.
\end{equation}
The invariants of the Weierstrass function are given by:
\begin{equation}
g_2=\frac{1}{12}+2Mr_mH^2-H^2r^2_m(1+H^2r^2_m),
\end{equation}
\begin{equation}
g_3=\frac{1}{216}+\frac{M^2H^2}{4}-\frac{Mr_mH^2}{3}+\frac{H^2r^2_m}{6}(1+
H^2r^2_m),
\end{equation}
from which one can calculate the discriminant, the roots etc. From eq. (4.61)
it is however already clear that the string starts with maximal size $r_m$ for
$\tau=0$, it then contracts and eventually collapses into the singularity
$r=0$ (taking for convenience $\tau_o=0$).
\vskip 6pt
\hspace*{-6mm}From Fig.2 and the above analysis we conclude that the circular
string motion is in fact very similar in the equatorial plane of the four
backgrounds of Minkowski space,
anti de Sitter space, Schwarzschild black hole and Schwarzschild anti de
Sitter space. In all these cases the string has a maximal size, and then
contracts towards $r=0$. Quantitatively there are of course differences but
qualitatively the motion from $r=r_m$ to $r=0$ is the same. This also includes
the $2+1$ dimensional BH-ADS spacetime when $J=0$: according to Fig.1b and
eq. (4.35), the $J=0$ circular string motion is qualitatively
similar to the four cases
described above.

This similarity can actually be pushed one step further by considering small
perturbations propagating around the circular strings, using the covariant
approach of Frolov and Larsen \cite{fro}. For a line element in the form (2.1)
the circular string is determined by eqs. (4.10)-(4.11):
\begin{equation}
\dot{t}=\frac{E}{a(r)},
\end{equation}
\begin{equation}
\dot{r}^2+r^2a(r)=E^2.
\end{equation}
We can introduce a normal vector $n^\mu$ perpendicular to the string
world-sheet ($x^\mu=t,r,\phi$):
\begin{equation}
g_{\mu\nu}\dot{x}^\mu n^\nu=g_{\mu\nu}x'^\mu n^\nu=0,\;\;\;\;g_{\mu\nu}n^\mu
n^\nu=1,
\end{equation}
explicitly given by:
\begin{equation}
n^\mu=\left( \frac{\dot{r}}{ra(r)},\;\frac{E}{r},\;0\right),
\end{equation}
and fulfilling the completeness relation:
\begin{equation}
g^{\mu\nu}=\frac{1}{r^2}(x'^\mu x'^\nu-\dot{x}^\mu\dot{x}^\nu)+n^\mu n^\nu.
\end{equation}
For circular strings in the equatorial plane of a higher dimensional spacetime
there will also be normal vectors in the directions perpendicular to the plane
of the string but they will not concern us here.  By defining the comoving
physical perturbations $\delta x$ to be the perturbations in the $n^\mu$
direction:
\begin{equation}
\delta x^\mu=\delta x\; n^\mu,
\end{equation}
it can be shown that \cite{all2}:
\begin{equation}
(\partial^2_\sigma-\partial^2_\tau)\delta x-\left( \frac{r}{2}\frac{da(r)}
{dr}+\frac{r^2}{2}
\frac{d^2a(r)}{dr^2}-\frac{2E^2}{r^2}\right)\delta x=0,
\end{equation}
to first order in the perturbation. To solve this equation one has to first
solve eq. (4.66) for $r(\tau)$. After Fourier expanding $\delta x$ we get
the Schr\"{o}dinger equation \cite{all2}:
\begin{equation}
\ddot{C}_n+\left( n^2+\frac{r}{2}\frac{da(r)}{dr}+\frac{r^2}{2}
\frac{d^2a(r)}{dr^2}-\frac{2E^2}{r^2}
\right) C_n=0,
\end{equation}
determining the Fourier components of the comoving perturbations.

The four non-rotating
spacetimes considered in this subsection are special cases of
$a(r)=1-2M/r+H^2r^2$, while the $J=0\;$-case of the $2+1$ BH-ADS spacetime
corresponds to $a(r)=\frac{r^2}{l^2}-M$. However, in all these cases it is
clear from eq. (4.72) that the comoving perturbations are regular except
near $r=0$, where the dominant term in the "potential" (4.72) is the
$E^2$-term. The $E^2$-term depends on $a(r)$ through the denominator $r^2$
that is obtained by solving eq. (4.66). In all the non-rotating cases
considered in this section it is however easy to see that for
$r\rightarrow 0$ we have $r\approx -E(\tau-\tau_o)$. It follows that not
only the unperturbed circular strings but also the comoving perturbations
around them behave in a qualitatively similar way in all the non-rotating
backgrounds considered in this section.
This should be contrasted with the analysis
of the string perturbations around the string center of mass (Section 3),
where we found qualitative differences between strings
in ordinary $3+1$ dimensional black hole spacetimes and strings in the $2+1$
dimensional BH-ADS spacetime. Notice however that in both approaches
we have only made the comparison for non-rotating spacetimes.
\section{Circular Strings with Unbounded Radius}
\setcounter{equation}{0}
In the previous section we concluded that the circular string motion is
qualitatively similar in the equatorial plane of spacetimes with line element:
\begin{equation}
ds^2=-a(r)dt^2+\frac{dr^2}{a(r)}+r^2d\theta^2+r^2\sin^2\theta d\phi^2,
\end{equation}
with:
\begin{equation}
a(r)=H^2r^2+1-\frac{2M}{r},
\end{equation}
and in the $2+1$ dimensional spacetime:
\begin{equation}
ds^2=-a(r)dt^2+\frac{dr^2}{a(r)}+r^2 d\phi^2,
\end{equation}
with:
\begin{equation}
a(r)=\frac{r^2}{l^2}-M.
\end{equation}
This similarity can be physically understood in the following way: The dynamics
of a circular string in a curved spacetime is determined by the string
tension and by the local gravity. The string tension will always try to
contract the circular string, while the local gravity can be either attractive
or repulsive. For the spacetimes (5.1) and (5.3) the local gravity is
proportional to the derivative of $a(r)$. It follows that in the cases
represented by eqs. (5.2) and (5.4) the local gravity is always positive
(considering only positive $M$ in eq. (5.2)), corresponding to attraction. So
in these cases both the string tension and the local gravity work in the
direction of contraction of the circular string, and therefore all
strings collapse to $r=0$.

The above argument
also suggests that we can find qualitatively different circular string
motions by considering spacetimes with regions of negative local gravity
(repulsion). In such spacetimes we can expect to find regions where the string
tension is dominating, regions where the negative local gravity is
dominating and regions where the two opposite effects are of the same order,
being a natural balance ensuring the existence of stationary
circular strings (such a solution actually exists in de Sitter space
\cite{fro,mic,veg4}).

We have already seen that in the rotating $(J\neq 0)$ $2+1$ dimensional
BH-ADS spacetime we can have string solutions qualitatively
different from the $J=0$ solutions, namely non-collapsing circular strings
(provided $J^2>4E^2$). The same happens in the equatorial plane of rotating
$(a\neq 0)$ $3+1$ dimensional spacetimes in the form (4.43). If
$a^2>E^2\Delta^2_o$ we see from eq. (4.46) that the potential is positive
infinite for $r\rightarrow 0$, so that no collapse into $r=0$ is possible.

A somewhat simpler example is provided by the Reissner-Nordstr\"{o}m black
hole, which has a region of negative local gravity inside the (outer)
horizon. Circular strings in Reissner-Nordstr\"{o}m background were
investigated in Ref.\cite{all2} and indeed non-collapsing solutions were
found.

It is also easy to find spacetimes with negative local gravity in the
asymptotic region $r\rightarrow\infty$. The simplest example is ordinary
de Sitter space, which in the static parametrization takes the form (5.1)
with $a(r)=1-H^2r^2$. In that case the potential (4.46) goes to minus infinity
$(V(r)\propto -r^4)$ for $r\rightarrow\infty$ and unbounded expanding
circular strings are found [6, 8]. These
solutions and the other types of circular
string solutions in de Sitter space were discussed in great detail in
Ref.\cite{veg4}, so we shall not say too much about it here. One of the most
important results in de Sitter space is the existence of
multi-string solutions [6-8]. It turns out that for certain
ranges of the integration constant $E$ (which is called $-\sqrt{b}/H$ in
Ref.\cite{veg4}) the internal world-sheet time $\tau$ is a multi (finite or
infinite) valued function of the cosmic time. This means that one single
world-sheet, where $\tau$ runs from $-\infty$ to $+\infty$, can describe
finitely or infinitely many different and independent strings in de Sitter
space. It is an interesting question whether this feature is also present
in other curved backgrounds.
Let us now consider briefly two other curved spacetimes in which
we find circular string solutions with unbounded $r,$ and multi-string
solutions.
\subsection{Circular Strings in Schwarzschild de Sitter Space}
Schwarzschild de Sitter space is in the form of eq. (5.1) with:
\begin{equation}
a(r)=1-\frac{2M}{r}-H^2r^2.
\end{equation}
It is a very interesting spacetime for circular strings for several reasons.
First of all it has regions of both positive $(da(r)/dr>0)$ local gravity and
regions of negative $(da(r)/dr<0)$ local gravity. Secondly, it is
asymptotically
de Sitter, so we expect to find features similar to the "pure" de Sitter case,
for instance the existence of multi-strings, as discussed above.

The mathematics of the circular strings is unfortunately going to be more
complicated here, as compared to the cases discussed in Section 4 and in
Ref.\cite{veg4}. From eq. (5.5) follows that Schwarzschild de Sitter spacetime
has two horizons (one horizon when equality) provided that $\sqrt{27}HM\leq 1$.
Explicitly they are given by $(r_+\geq r_-)$:
\begin{equation}
r_-=M\left( -\frac{1-i\sqrt{3}}{2^{2/3}Z(HM)}-\frac{(1+i\sqrt{3})Z(HM)}{6H^2
M^2 2^{1/3}}\right),
\end{equation}
\begin{equation}
r_+=M\left( \frac{2^{1/3}}{Z(HM)}+\frac{Z(HM)}{3H^2 M^2 2^{1/3}}\right),
\end{equation}
where we introduced the notation:
\begin{equation}
Z(HM)\equiv[-54H^4 M^4+\sqrt{-108H^6 M^6(1-27H^2 M^2)}\;]^{1/3},
\end{equation}
and we only consider the region $HM\in\;]0,\;1/\sqrt{27}]$. The circular
string potential (4.46) for $a=0$ and $\Lambda\equiv 3H^2>0$ takes the form:
\begin{equation}
V(r)=-H^2r^4+r^2-2Mr-E^2,
\end{equation}
so that:
\begin{equation}
V(0)\;=\;V(r_+)\;=\;V(r_-)\;=\;-E^2,\;\;\;V(r)\propto
-r^4\;\;\;\mbox{for}\;\;\;r>>(M,H^{-1},E).
\end{equation}
The potential has a local minimum between $r=0$ and the inner horizon, and
a local maximum at $r=r_o$ between the two horizons:
\begin{equation}
\frac{dV(r)}{dr}\mid_{r=r_o}=0,\;\;\;\frac{d^2V(r)}{dr^2}\mid_{r=r_o}<0,
\end{equation}
where:
\begin{equation}
r_o=M\left( \frac{2^{1/3}}{W(HM)}+\frac{W(HM)}{6H^2 M^2 2^{1/3}}\right),
\end{equation}
and:
\begin{equation}
W(HM)\equiv[-108H^4M^4+\sqrt{-432H^6 M^6(2-27H^2M^2)}\;]^{1/3}.
\end{equation}
Since the string dynamics takes place at the $r$-axis in a $(r,V(r))$ diagram,
it is important to know exactly the shape of the potential. If $V(r_o)>0$ the
potential has two zeroes between the two horizons, and it will act effectively
as a barrier, see Fig.3. On the other hand, if $V(r_o)<0$ there is no barrier
and nothing can prevent a contracting string from collapsing into the
singularity $r=0$. One finds:
\begin{equation}
V(r_o)>0\;\Leftrightarrow \frac{E^2}{M^2}<-H^2 M^2(\frac{r_o}{M})^4+
(\frac{r_o}{M})^2-2(\frac{r_o}{M}),
\end{equation}
where the right hand side of the (second) inequality depends on $HM$ only. The
inequality (5.14) also provides the critical value of $(E/M)$ in terms of
$HM$, when the potential equals zero for $r=r_o$.

The mathematical solution of eqs. (4.45) for $r(\tau),$
determining
the invariant string size
as a function of $\tau$, can be formally obtained from the Schwarzschild
anti de Sitter case, discussed in Subsection 4.2, by simply changing the sign
of $H^2$ in eqs. (4.59)-(4.64). The parameter $r_m$ defined in eq. (4.60)
however, can no longer
be interpreted as the maximal string size. In the present
case we can just take $r_m$ to be any of the complex roots of the quartic
equation (4.60) (with $H^2$ replaced by $-H^2$), and then also the (complex)
constant
$\tau_o$ introduced in eq. (4.61) must be carefully chosen to obtain a real
$r(\tau)$ for real $\tau$.

In the present paper we shall not go into a complete analysis of the various
types of solutions. We will restrict ourselves by considering only the
"degenerate" case $V(r_o)=0$, where the Weierstrass function reduces to an
elementary function. In that case it is tempting to take $r_o=r_m$, but then
the constants $d_1$ and $d_2$, defined in eq. (4.62), diverge, so we have
to take one of the other roots of eq. (4.60). After a little algebra we
find:
\begin{equation}
r(\tau)=r_o+\frac{4(6H^2r^2_o-1)e^{-\sqrt{6H^2r^2_o-1}\;(\tau-\tau_o)}}
{H(e^{-\sqrt{6H^2r^2_o-1}\;(\tau-\tau_o)}-4H^2r^2_o)^2-4H(6H^2r^2_o-1)}.
\end{equation}
It is clear from the potential (Fig.3) that there are two qualitatively
different types of solutions, which we shall call $r_+(\tau)$ and
$r_-(\tau)$.

For:
\begin{equation}
\tau_o=\frac{1}{\sqrt{6H^2r^2_o-1}}\log\frac{2-8H^2r^2_o+2\sqrt{6H^2r^2_o-1}
\sqrt{3H^2r^2_o-1}}{Hr_o},
\end{equation}
we have the solution $r_+(\tau)$ with the properties:
\begin{equation}
r_+(-\infty)=r_o,\;\;\;r_+(0)=0.
\end{equation}
This circular string starts with its maximal size $r_o$ at $\tau=-\infty$,
passes the inner horizon and falls into the singularity $r=0$ at $\tau=0$.

The choice:
\begin{equation}
\tau_o=\frac{1}{\sqrt{6H^2r^2_o-1}}\log(4Hr_o+2\sqrt{6H^2r^2_o-1}),
\end{equation}
leads to a different type of solution, $r_-(\tau)$, with:
\begin{equation}
r_-(-\infty)=r_o,\;\;\;r_-(0)=\infty,\;\;\;r_-(\infty)=r_o.
\end{equation}
This solution is very similar to the multi-string solution discussed in
Refs.\cite{mic,veg4} for strings in de Sitter space.
Each of the two world-sheet time intervals $\tau\in\;
]-\infty,\;0]$ and $\tau\in[0,\;\infty\;[$ corresponds to the physical
time interval $\;]-\infty,\;\infty[\;$, that is, the world-sheet time $\tau$ is
a two-valued function of the physical time, and eqs. (5.15), (5.18)-(5.19)
describe a multi (two) string solution.

More generally, from the potential (Fig.3) and the similarity between the
Schwarzschild de Sitter and the
"pure" de Sitter spacetimes outside the horizon,
we can expect multi-strings for any values of the parameters $(M,H,E).$
\subsection{Circular Strings in the Black String Background}
As our final example of circular strings in curved spacetimes, we consider
the $2+1$ dimensional black string of Horne and Horowitz \cite{hor2}. For
our purposes it is most convenient to use the stationary (but non-static)
parametrization (2.13), which in the general form (4.1) is:
\begin{equation}
\tilde{g}_{tt}=M-\frac{J^2}{4r^2},\;\;\;\tilde{g}_{rr}=(\frac{r^2}{l^2}-M
+\frac{J^2}{4r^2})^{-1},\;\;\;\tilde{g}_{\phi\phi}=\frac{1}{r^2},\;\;\;
\tilde{g}_{t\phi}=\frac{1}{l},
\end{equation}
where the tilde reminds us that the black string is dual to the $2+1$
dimensional BH-ADS spacetime (2.6). From the analysis of Section 4 we can
then write down the equations (4.9)-(4.11), determining the circular string
motion in the background (5.20):
\begin{equation}
\dot{r}^2+\tilde{V}(r)=0,\;\;\;\;\dot{t}=\frac{-E}{M-\frac{J^2}{4r^2}-
\frac{r^2}{l^2}},\;\;\;\;\dot{f}=\frac{E}{\frac{l}{r^2}(M-\frac{J^2}{4r^2})-
\frac{1}{l}}.
\end{equation}
For the $r$-equation we find a potential
which is dual to the potential for the $2+1$ dimensional BH-ADS spacetime, in
the sense that:
\begin{equation}
\tilde{V}(r)=\tilde{g}^{rr}(E^2\tilde{g}^{tt}+\tilde{g}_{\phi\phi}),
\end{equation}
(compare with eqs. (4.11), (4.15)), and given explicitly by:
\begin{equation}
\tilde{V}(r)=\frac{J^2}{4r^4}-\frac{M}{r^2}+\frac{1}{l^2}-E^2.
\end{equation}
It follows that:
\begin{equation}
\tilde{V}(r^2_\pm)=-E^2,\;\;\;\tilde{V}(\infty)=\frac{1}{l^2}-E^2,
\end{equation}
where $r_\pm$ are the two horizons (2.7), unchanged by the duality
transformation. For $r\rightarrow 0$ the potential is positive infinite for
$J\neq 0$, and negative infinite for $J=0\;$ (taking $M$ positive).

One effect of the duality transformation has been to change the asymptotic
behaviour of the circular string potential, compare with eq. (4.15). In the
case that
$1/(l^2)-E^2<0$ the potential approaches a negative constant value, which
gives rise to string solutions of unbounded $r$. In this sense we will find
solutions of completely different type than the solutions discussed in
Subsection 4.1, that were always bounded. Another effect of the duality
transformation is however to change the expression for the invariant string
size, defined in eq. (4.14):
\begin{equation}
\tilde{S}(\tau)=\sqrt{\tilde{g}_{\phi\phi}}=\frac{1}{r(\tau)}.
\end{equation}
Writing eqs. (5.21) in terms of the invariant string size yields:
\begin{equation}
\dot{\tilde{S}}^2+\tilde{V}(\tilde{S})=0,\;\;\;\;\dot{t}
=\frac{-E}{M-\frac{J^2}{4}\tilde{S}^2-
(l\tilde{S})^{-2}},\;\;\;\;\dot{f}=\frac{E}{l(M-\frac{J^2}{4}\tilde{S}^2)
\tilde{S}^2-\frac{1}{l}},
\end{equation}
where:
\begin{equation}
\tilde{V}(\tilde{S})=\tilde{S}^4(\frac{J^2}
{4}\tilde{S}^4-M\tilde{S}^2+\frac{1}{l^2}-E^2).
\end{equation}
For non-zero $J$, which means non-zero charge for the black string \cite{hor2},
we have only bounded configurations (finite $\tilde{S}$), while
for $J=0$ and $M>0,$ unbounded configurations
will exist as well. The potential (5.27) is shown in Fig.4 for various values
of the parameters. In the simplest case ($J=0,\;E^2l^2=1$) eqs.
(5.26)-(5.27) are solved by:
\begin{equation}
\tilde{S}(\tau)=\frac{1}{\sqrt{\mid 2\sqrt{M}\tau\;\mid}},
\end{equation}
\begin{equation}
t(\tau)=\frac{1}{2E\sqrt{M}}\log\mid\frac{2E^2}{\sqrt{M}}\tau-1\mid,
\end{equation}
\begin{equation}
\phi(\tau,\sigma)=\sigma-\tau-\frac{\sqrt{M}}{2E^2}\log\mid\frac{2E^2}
{\sqrt{M}}\tau-1\mid,
\end{equation}
so that:
\begin{eqnarray}
&\tilde{S}(-\infty)=0,\;\;\;\;\tilde{S}(0)=\infty,\;\;\;\;\tilde{S}(+\infty)=
0,&\nonumber\\
&t(-\infty)=\infty,\;\;\;\;t(0)=0,\;\;\;\;t(\infty)=\infty,&
\end{eqnarray}
This solution is somewhat similar to the two-string solution found in de
Sitter space (Refs.\cite{mic,veg4})
and in Schwarzschild de Sitter space (Subsection
5.1), and indicates that multi-string solutions are a generic feature of
the black string spacetime too.

In the general case the
mathematical solution of eqs. (5.26)-(5.27), or alternatively
of eqs. (5.21)-(5.22),
can be obtained in terms of elliptic functions (elementary functions for
$J=0$). This solution
and its physical interpretation, are to be discussed
elsewhere.
\section{Conclusion}
We have studied the string propagation in the 2+1 BH-ADS and black string
backgrounds. We found the first and second order perturbations around the
string center of mass as well as the mass formula, and compared with the
ordinary black hole ADS spacetime. We found the exact general evolution
of circular strings in all these backgrounds, and in the black hole de Sitter
spacetime: in all these cases the solutions were expressed closely and
completely in terms of either elementary or elliptic functions.
The physical properties of the string
motion in these backgrounds have been discussed.
A summary of the main features and conclusions of our paper is given in
Tables I and II.

\newpage
\begin{center}
{\bf Figure Captions}
\end{center}
\vskip 12pt
Fig.1. The potential $V(r)$ eq. (4.15) for a circular string in the
$2+1$ dimensional black hole anti de Sitter (BH-ADS)
spacetime. In (a) we have $J^2>4E^2$ and a barrier
between the inner horizon and $r=0$, while (b) represents a case where
$J^2<4E^2$ and a string will always fall into $r=0$. In the cases shown, the
values of the various parameters are: $M=1$, $J=l/\sqrt{2}$ as well as
$E=0.1\;$(case (a)) and
$E=0.5\;$(case (b)). The static limit is $r_{\mbox{erg}}=l$.
\vskip 36pt
\hspace*{-6mm}Fig.2. The potential $V(r)$ eq. (4.46) for
a circular string in the
equatorial plane of the four $3+1$ dimensional spacetimes: (a) Minkowski
space (MIN), (b) Schwarzschild black hole (S), (c) anti de Sitter space (ADS),
and (d) Schwarz- schild
anti de Sitter space (S-ADS). The potentials are plottet
for fixed $E$ and we notice the following general relations between the
maximal string radii $r_m$: $r^{S}_m>r^{MIN}_m>r^{ADS}_m,\;\;\;r^{S}_m>
r^{S-ADS}_m>r^{ADS}_m$ and $r^{MIN}_m>r^{S-ADS}_m\;\Leftrightarrow\;
H^2 E^3>2M$.
\vskip 36pt
\hspace*{-6mm}Fig.3. The potential $V(r)$ eq. (5.9) for a
circular string in the ordinary
Schwarzschild de Sitter spacetime (SDS). For $V(r_o)\geq0$ there
are qualitatively
different types of solutions, since $V(r)$ acts as a barrier.
\vskip 36pt
\hspace*{-6mm}Fig.4. The potential $\tilde{V}(\tilde{S})$ eq. (5.27) for
a circular string in the
black string background. For $M>0$ and $M^2l^2-J^2\geq 0$ (the latter is the
condition for the existence of horizons) there are four qualitatively
different cases: (a) $J=0,\;E^2l^2<1;\;$ (b) $J=0,\;E^2l^2\geq 1;\;$
(c) $J\neq 0,\;E^2l^2\geq 1;\;$ (d) $J\neq 0,\;E^2l^2<1.$
\newpage
\begin{center}
{\bf Table Captions}
\end{center}
\vskip 12pt
Table I. String motion described by the string perturbation series approach in
the 2+1 BH-ADS, ordinary black hole-ADS, de Sitter (DS) and
black string backgrounds.
Notice the difference between the string motion in the 2+1 BH-ADS and in
the other spacetimes, while the strings in the black string background
behave similarly as in the ordinary black hole backgrounds.
\vskip 36pt
\hspace*{-6mm}Table II. Circular exact string solutions in the indicated
backgrounds. $S(\tau)$ is the invariant string size. The motion is exactly
and completely solved in terms of elliptic
functions; ($\wp(\tau-\tau_0)$ stands for the Weierstrass elliptic
function which reduces to elementary functions for zero cosmological constant
or for particular combinations of the spacetime parameters). The properties
of the potential $V(r)\;\;(\tilde{V}(r)$ in the dual black string
background) determine general properties of the string dynamics. Unstable
strings (expanding with unbounded size) and multi-string solutions are
present for potentials unbounded from below for $r\rightarrow\infty,$ while
bounded strings (and no multi-string solutions)
correspond to $V(\infty)=+\infty.$
When $V(0)<0,$ strings can collapse into a point.
\end{document}